\documentclass[%
 aip,
 jmp,%
 amsmath,amssymb,
 reprint,%
]{revtex4-1}
\usepackage{hyperref}
\usepackage{graphicx}
\usepackage{dcolumn}
\usepackage{bm}
\usepackage[mathlines]{lineno}
\usepackage[english]{babel}
\usepackage{amsmath}
\usepackage{graphics}
\usepackage{setspace}
\usepackage[colorinlistoftodos]{todonotes}
\usepackage{wrapfig}
\usepackage{float}
\usepackage{balance}
\usepackage{array}
\usepackage{mathtools}
\usepackage{multirow}
\usepackage[normalem]{ulem}
\begin{document}
\preprint{AIP/123-QED}
\title[The Advanced LIGO Photon Calibrators]{The Advanced LIGO Photon Calibrators}
\author{S. Karki}%
\email{skarki@uoregon.edu.}
\affiliation{University of Oregon, Eugene, OR, 97403, USA}
\affiliation{LIGO Hanford Observatory, Richland, WA, 99352, USA}%
\author{D. Tuyenbayev}%
\affiliation{LIGO Hanford Observatory, Richland, WA, 99352, USA}
\affiliation{University of Texas Rio Grande Valley, Brownsville, TX, 78520, USA}
\author{S. Kandhasamy}%
\affiliation{University of Missisipi, Oxford, MS, 38677, USA}
\affiliation{LIGO Livingston Observatory, Livingston, LA, 70754, USA }
\author{B. P. Abbott}
\affiliation{California Institute of Technology, Pasadena, CA, 91125, USA}
\author{T. D. Abbott}%
\affiliation{Louisiana State University, Baton Rouge, LA 70803, USA}
\author{E. H. Anders}%
\affiliation{LIGO Hanford Observatory, Richland, WA, 99352, USA}
\author{J. Berliner}%
\affiliation{LIGO Hanford Observatory, Richland, WA, 99352, USA}
\author{J. Betzwieser}%
\affiliation{LIGO Livingston Observatory, Livingston, LA, 70754, USA }
\author{H. P. Daveloza}%
\affiliation{University of Texas Rio Grande Valley, Brownsville, TX, 78520, USA}
\author{C. Cahillane}%
\affiliation{California Institute of Technology, Pasadena, CA, 91125, USA}
\author{L. Canete}%
\affiliation{LIGO Hanford Observatory, Richland, WA, 99352, USA}
\author{C. Conley}%
\affiliation{LIGO Hanford Observatory, Richland, WA, 99352, USA}
\affiliation{California Institute of Technology, Pasadena, CA, 91125, USA}
\author{J. R. Gleason}%
\affiliation{University of Florida, Gainesville, FL, 32611, USA}
\author{E. Goetz}%
\affiliation{LIGO Hanford Observatory, Richland, WA, 99352, USA}
\affiliation{California Institute of Technology, Pasadena, CA, 91125, USA}
\author{J. S. Kissel}%
\affiliation{LIGO Hanford Observatory, Richland, WA, 99352, USA}
\author{K. Izumi}%
\affiliation{LIGO Hanford Observatory, Richland, WA, 99352, USA}
\author{G. Mendell}%
\affiliation{LIGO Hanford Observatory, Richland, WA, 99352, USA}
\author{V. Quetschke}%
\affiliation{University of Texas Rio Grande Valley, Brownsville, TX, 78520, USA}
\author{M. Rodruck}%
\affiliation{LIGO Hanford Observatory, Richland, WA, 99352, USA}
\author{S. Sachdev}%
\affiliation{California Institute of Technology, Pasadena, CA, 91125, USA}
\author{T. Sadecki}%
\affiliation{LIGO Hanford Observatory, Richland, WA, 99352, USA}
\author{P. B. Schwinberg}
\affiliation{LIGO Hanford Observatory, Richland, WA, 99352, USA}
\author{A. Sottile}
\affiliation{LIGO Hanford Observatory, Richland, WA, 99352, USA}
\affiliation{University of Pisa, Pisa, Italy}
\author{M. Wade}%
\affiliation{Kenyon College, Gambier, OH 43022, USA}
\author{A. J. Weinstein}%
\affiliation{California Institute of Technology, Pasadena, CA, 91125, USA}
\author{M. West}%
\affiliation{LIGO Hanford Observatory, Richland, WA, 99352, USA}
\affiliation{Syracuse University, Syracuse, NY, 13244, USA}
\author{R. L. Savage}
\email{richard.savage@ligo.org}
\affiliation{LIGO Hanford Observatory, Richland, WA, 99352, USA}%
\date{\today}%
\renewcommand{\arraystretch}{1.4}
\newcolumntype{C}[1]{>{\centering\arraybackslash}m{#1}}
\begin{abstract}
The two interferometers of the Laser Interferometry Gravitaional-wave Observatory (LIGO) recently detected gravitational waves from the mergers of binary black hole systems. Accurate calibration of the output of these detectors was crucial for the observation of these events, and the extraction of parameters of the sources. The principal tools used to calibrate the responses of the second-generation (Advanced) LIGO detectors to gravitational waves are systems based on radiation pressure and referred to as Photon Calibrators. These systems, which were completely redesigned for Advanced LIGO, include several significant upgrades that enable them to meet the calibration requirements of second-generation gravitational wave detectors in the new era of gravitational-wave astronomy.  We report on the design, implementation, and operation of these Advanced LIGO Photon Calibrators that are currently providing fiducial displacements on the order of $10^{-18}$~m/$\sqrt{\textrm{Hz}}$ with accuracy and precision of better than 1\,\%.
\end{abstract}


\maketitle

\section{Introduction}\label{Introduction}
On September 14, 2015, 100 years after the first prediction of the existence of gravitational waves, the Advanced Laser Interferometer Gravitational-wave Observatory (LIGO) detected the gravitational-wave signals emitted by the merger of a binary black hole system, GW150914.~\cite{detection} Additional signals have been detected since then.~\cite{detection2,P1600088} These observations have initiated the era of gravitational wave astronomy. Accurately reconstructing the gravitational wave signals requires precise and accurate calibration of the responses of the detectors to variations in the relative lengths of the 4-km-long interferometer arms.~\cite{P1500248} Extracting the parameters of the events that generated the waves also imposes stringent requirements on detector calibration.~\cite{P1500218}  The estimated required calibration accuracy for LIGO's initial detection phase was on the order of 5\,\%, while the requirements for making precision measurements of source parameters are on the order of 0.5\,\%.~\cite{Lindblom}

\begin{figure}
\includegraphics[width = .5\textwidth]{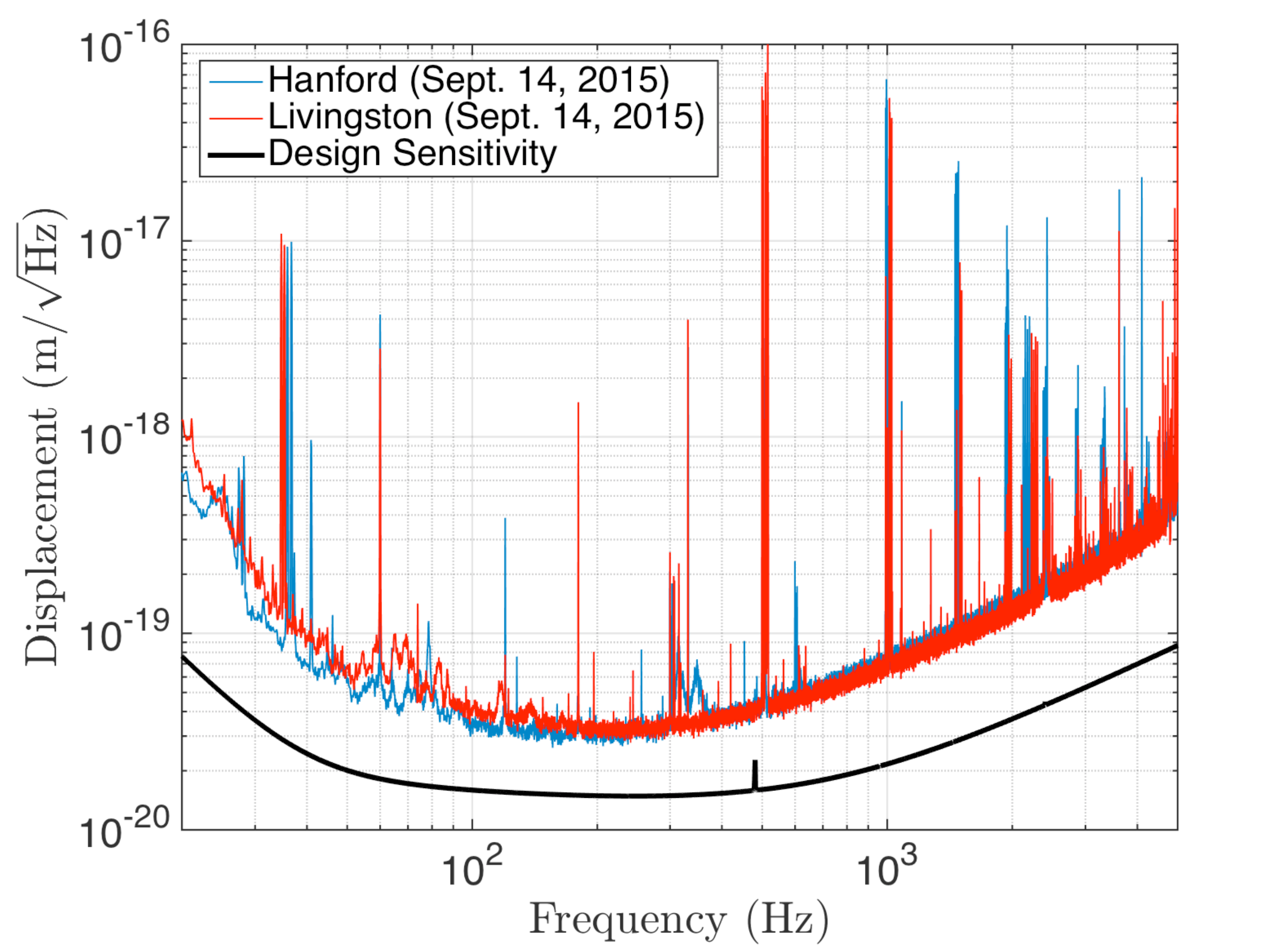}
\caption{Relative displacement sensitivity of the Hanford (red) and Livingston (blue) interferometers in Sept., 2015.  The black curve is the design sensitivity.  The sharp features in the spectra are from calibration lines (37 Hz, 332 Hz, 1.1 kHz), AC power lines (60 Hz and harmonics), and mirror suspension fiber {\em violin-mode} resonances (500 Hz and harmonics).}
\label{fig:SensCurves}
\end{figure}

The Advanced LIGO detectors located in Richland, Washington, and Livingston, Louisiana are variants of Michelson laser interferometers with enhancements aimed at increasing their sensitivity to differential length variations, which are the signature of passing gravitational waves.~\cite{detectorpaper} These enhancements include 4-km-long Fabry-Perot resonators in the arms,  power recycling, and resonant sideband extraction.~\cite{mizuno} The displacement sensitivity during the GW150914 event and the Advanced LIGO design sensitivity are shown in Fig.~\ref{fig:SensCurves}.~\cite{P1500260}
The peak sensitivity of about $3\times 10^{-20}$~m/$\sqrt{\textrm{Hz}}$ was achieved for differential length variations at frequencies near 200~Hz.  To achieve this level of displacement-equivalent background noise, isolation of the arm cavity mirrors (serving as test masses for gravitational waves) from ground motion requires sophisticated vibration isolation systems.~\cite{sei} The 40\,kg mirrors are suspended from cascaded quadruple pendulums and controlled by contact-free electrostatic actuators.~\cite{suspension}  Calibration of the differential length responses of the interferometers requires inducing fiducial periodic length variations at the level of $10^{-15}$ to $10^{-18}$~m/$\sqrt{\textrm{Hz}}$ over a range of frequencies from a few hertz to several kHz.    

Photon Calibrators (Pcals) are the primary calibration tool for the Advanced LIGO detectors. Earlier versions have been tested on various interferometers~\cite{virgopcal,Glasgowpcal,GEO600pcal} and they have evolved significantly within LIGO over the past ten years.~\cite{P080118} These systems operate during observing periods, providing continuous calibration information while the detectors are in their most sensitive configuration -- a distinct advantage over other calibration techniques.~\cite{iLIGocal}

Pcals rely on photon radiation pressure from auxiliary, power-modulated laser beams reflecting from a test mass to apply periodic forces via the recoil of photons. The periodic force on the mirror, directly proportional to the amplitude of the laser power modulation, results in modulation of the position of the mirror and therefore the length of the arm cavity. Measuring the modulated laser power reflecting from the mirror with the required accuracy is one of the principal challenges for Pcal systems.

The fiducial length modulation, $x(f)$, induced by modulated Pcal power, $P(f)$, is given by~\cite{P080118}
\begin{equation}
x(f) = \frac{2 \cos\theta}{c}\left[1+\frac{M}{I}(\vec{a}\cdot\vec{b})\right]S(f)\, P(f)
\label{eq:pcaldisp}
\end{equation}
where $\theta$ is the angle of incidence of the Pcal beams on the test mass surface, $c$ is the speed of light, $M$ is the mass of the mirror, $I$ is its rotational moment of inertia, $\vec{a}$ and $\vec{b}$ are displacement vectors from the center of the test mass for the Pcal center of force and the interferometer beam, respectively, and $S(f)$ is the force-to-length transfer function of the suspended test mass. For Advanced LIGO mirror suspensions at frequencies above 20 Hz, $S(f)$ is well approximated by the free-mass response, $S(f)\approx-1/[M(2 \pi f)^{2}]$.~\cite{P1500248} The term $ (\vec{a}\cdot\vec{b})M/I$, accounts for unintended effective length changes resulting from rotation of the test mass induced by applied Pcal forces. 

These Pcal forces can also induce both {\em{local}}~\cite{P070074} and {\em{bulk}}~\cite{P1100166} elastic deformations of the test mass, compromising the accuracy of the calibration. To minimize the impact of these deformations, the Photon calibrators use two beams displaced symmetrically from the center of the face of the mirror and precisely positioned to reduce excitation of the natural vibrational modes of the mirror substrate.

Furthermore, because the Pcal forces are applied directly to the test masses, minimizing introduction of displacement  noise at frequencies other than the intended modulation frequencies is critical. The Pcals employ feedback control loops that ensure that the modulated power output match the requested waveform,  reducing the free-running relative power noise of the laser as well as harmonics of the modulation.

Four Advanced LIGO Pcal systems have been installed and are operating continuously, two at each LIGO observatory, one for each test mass at the ends of the interferometer arms. They are providing the required fiducial displacements with accuracy of better than one percent.

The remainder of this paper is organized as follows: in Sec.~\ref{Hardware} we give a detailed description of the instrument hardware and its capabilities; in Sec.~\ref{Calibration} absolute calibration of the laser power sensors is described; in Sec.~\ref{Uncertainties} uncertainties associated with Pcal-induced displacements are described; in Sec.~\ref{Applications} we discuss how Pcals are used in Advanced LIGO detectors to obtain the required calibration accuracy. Finally, conclusions are presented in Sec.~\ref{Conclusions}.

\section{Instrument Description}\label{Hardware}
\begin{figure*}
\includegraphics[width = 0.9\textwidth]{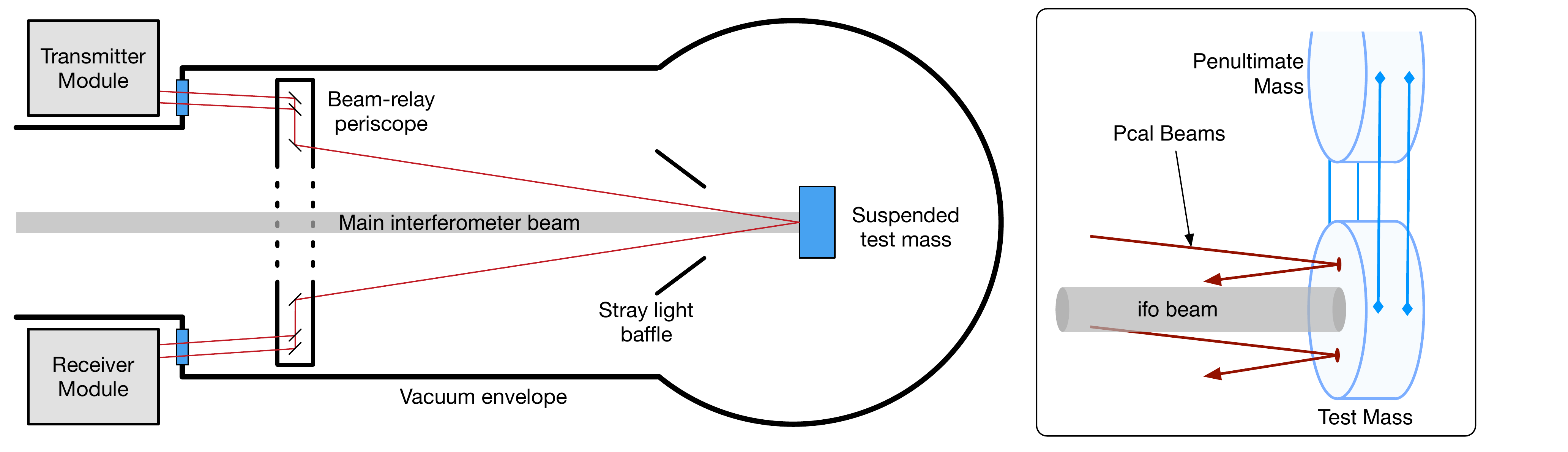}
\caption{Schematic diagram of an Advanced LIGO photon calibrator in plan view (left). The transmitter module contains the laser, power modulator, and beam conditioning optics.  The in-vacuum periscope structure relays the input beams to avoid occlusion by the stray-light baffling and to impinge on the end test mass at the desired locations.  It also relays the reflected beams to a power sensor mounted inside the receiver module. Schematic diagram of beams impinging on a suspended test mass surface (right).  The Pcal beams are displaced symmetrically above and below the center of the optic.  The main interferometer beam is nominally centered on the surface.}
\label{fig:pcal_layout}
\end{figure*}
Using the Advanced LIGO Pcals as the primary calibration tool increases demands for reliability and system performance.  To improve reliability, two Pcal systems are installed on each Advanced LIGO interferometer.  One Pcal system is sufficient for simultaneously injecting the several required displacement modulations at different frequencies (this is discussed in more detail in Sec.~\ref{Applications}). The other system serves as a backup and can be used to inject simulated gravitational-wave signals to test detection pipelines.~\cite{P1500269}

\begin{figure}
\includegraphics[width = .48\textwidth]{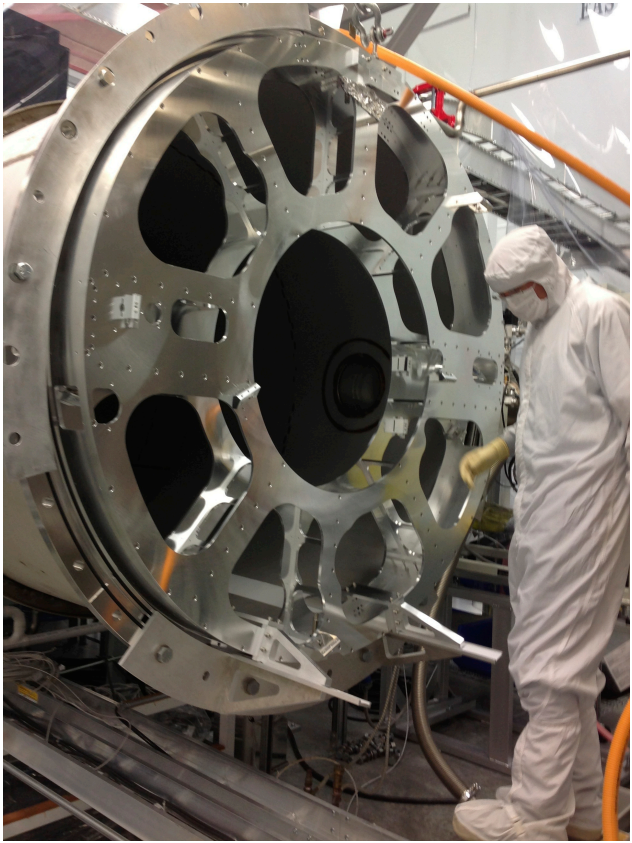}
\caption{The periscope structure that supports the relay optics that provide optical paths for the Pcal beams and the beam localization camera system being installed during the Advanced LIGO upgrade.}
\label{fig:periscope}
\end{figure}

A schematic diagram of an Advanced LIGO Pcal system is shown in Fig.~\ref{fig:pcal_layout}.
The transmitter and receiver modules, which are described in detail in Sec.~\ref{IIA}, are located outside the vacuum envelope.  The two beams from the transmitter module enter the vacuum enclosure through optical-quality, super-polished windows with low-loss ion beam sputtered anti-reflection coatings.  The specified transmissivity is greater than 99.6\,\%. These windows are an important element of the photon calibrators because optical losses are a significant component of the overall system uncertainty, as will be discussed in Sec.~\ref{Uncertainties}.  Each of the horizontally-displaced input beams is relayed by mirrors mounted to a periscope structure located inside the vacuum envelope to reduce the angle of incidence on the end test mass and thus avoid occlusion by stray light baffles.  Installation of a periscope frame into the vacuum envelope during the Advanced LIGO upgrade is shown in Fig.~\ref{fig:periscope}.
The beams from the in-vacuum periscope impinge on the test mass at 8.75 deg., displaced vertically by approximately 111.6 mm above and below the center of the mirror (see Fig.~\ref{fig:pcal_layout}).

The power reflectivity of the end test mass, measured in-situ with the Pcal beams, is 0.9979 \,$\pm$\,0.0010.~\cite{DarkhanAlog} The reflected beams are relayed by a second set of mirrors mounted to the in-vacuum periscope structure and exit the vacuum enclosure through an identical vacuum window. These beams enter the receiver module and are directed by a pair of mirrors to a power sensor mounted inside the receiver module. Capturing the light reflected from the test mass is an important upgrade because it enables tracking changes in the overall optical efficiency of the Pcal system.  Furthermore, it enables measurement of the full power, rather than just a sample of the power that is subject to changes in the reflectivity of the beam sampling optic.

Reducing calibration uncertainties requires higher signal-to-noise ratios (SNRs) for the fiducial length modulations, which requires increased laser power and thus Advanced LIGO Pcals have 2-watt lasers, four times the initial LIGO laser power. However, because they operate continuously at high SNR levels during observation runs, broadband laser power noise as well as harmonics of the injected modulations resulting from non-linearities in the modulation process must be minimized. To meet the Advanced LIGO requirement that unwanted noise injected by the Pcals be at least a factor of ten below the noise floor of the detector~\cite{T1100068}, a high-bandwidth feedback control servo known as the {\em Optical Follower Servo} (OFS) has been implemented.~\cite{T130442} The features and performance of this servo are described in detail in Sec.~\ref{IIB}.

Another important aspect of the performance of the Pcal systems is the locations of the Pcal beam spots on the test mass surface.  To minimize calibration errors resulting from local deformations of the test mass surface that are sensed by the interferometer beam, the Pcals use two beams with equal powers and displaced from the center of the mirror surface (the nominal location for the interferometer beam).  To minimize inducing rotation of the test mass, the two Pcal beams are displaced symmetrically about the center of the face of the mirror.  To minimize the impact of bulk elastic deformation of the mirror, the beams are located on the nodal circle of the {\em drumhead} natural vibrational mode. While this minimizes the deformation of the mirror in the drumhead mode shape, it efficiently deforms the mirror in the the lower-resonant-frequency {\em butterfly} mode shape. However, when the interferometer laser beam is centered on the mirror the butterfly mode integrates to zero over the central circular region.  Thus, the errors induced by excitation of this mode shape are minimal for small displacements of the interferometer beam from center. In order to determine and adjust the positions of the Pcal beams, a beam localization camera system has been implemented for Advanced LIGO. It is described in detail in Sec.~\ref{IIC}.
\subsection{Transmitter and Receiver Modules}\label{IIA}
\begin{figure}
\includegraphics[width = .5\textwidth]{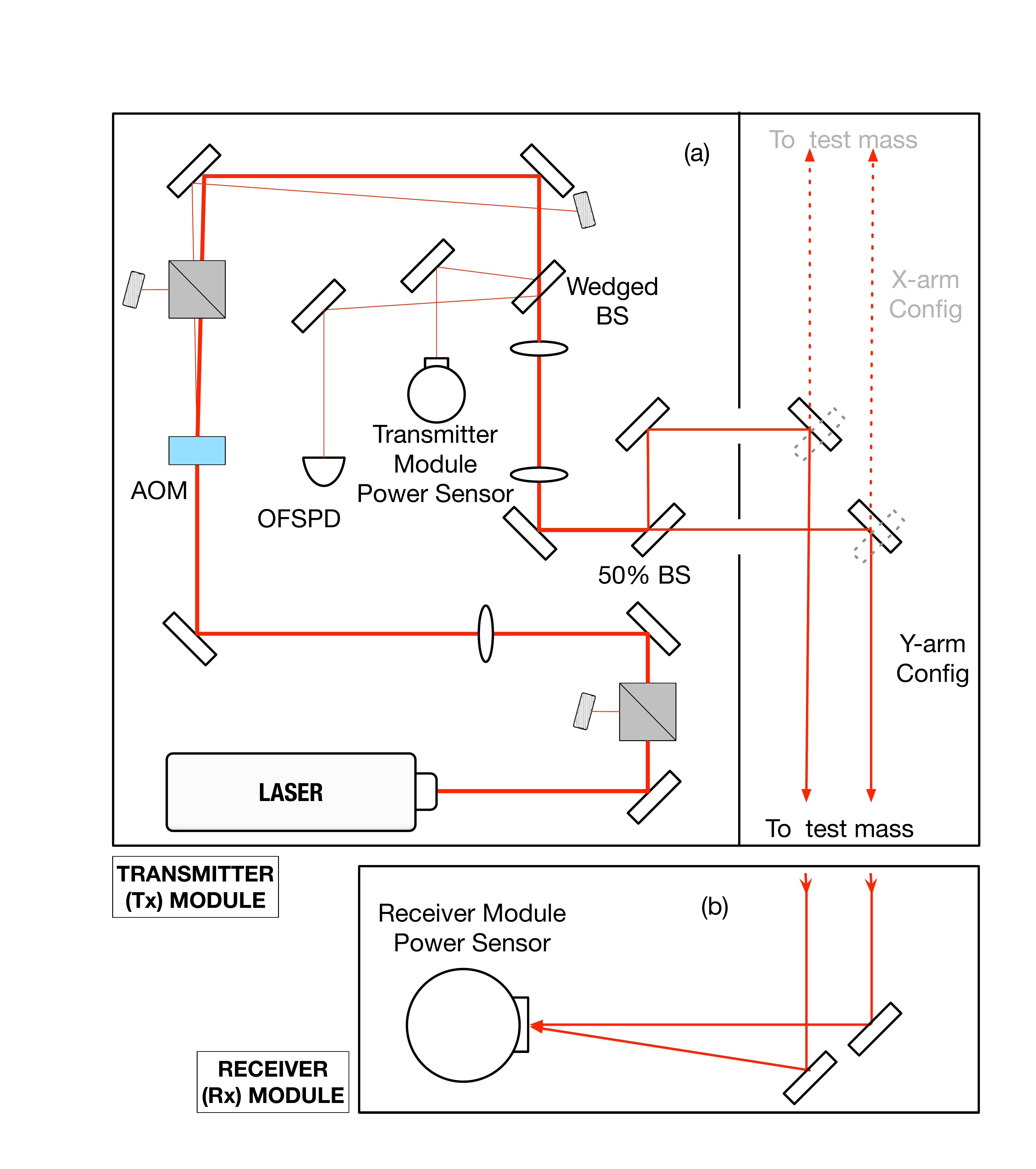}
\caption{(a)Schematic diagram of the optical layout of the transmitter module. The first-order diffracted beam from the acousto-optic modulator (AOM) is directed through an uncoated wedged beamsplitter at Brewster's angle to generate the sample beams for the two photodetectors. The transmitted beam is divided into two beams of equal power and directed toward the test mass located inside the vacuum envelope. (b) Schematic diagram of the optical layout of the receiver module.  The 4 in.\ diameter integrating sphere captures all of the Pcal light reflected from the test mass and transmitted through the output vacuum window.}
\label{fig:Txmodule_layout}
\end{figure}

The optical layout of the transmitter module is shown in Fig.~\ref{fig:Txmodule_layout}~(a). 
It houses a 2-watt Nd:YLF laser operating at 1047~nm.  This wavelength is close enough to the 1064~nm wavelength of the main interferometer laser to ensure high reflectivity from the test mass mirror coating, but far enough away to ensure that scattered Pcal light does not compromise interferometer operations. The horizontally-polarized output beam is focused into an acousto-optic modulator operating in the Littrow configuration that diffracts a fraction of the light in response to a control signal that changes the amplitude of the 80 MHz radio-frequency drive signal.  The maximum diffraction efficiency is approximately 80\,\%.  The non-diffracted beam is dumped and the first-order diffracted beam is directed through an uncoated wedge beamsplitter oriented near Brewster's angle that generates the sample beams used for two photodetectors.  The first sample beam is directed into a 2 in.\ diameter integrating sphere with an InGaAs photodetector. This system monitors the power directed into the vacuum system.  The second sample beam is directed to a similar photodetector (without the integrating sphere) that is the sensor for the {\em Optical Follower Servo} described in Sec.~\ref{IIB}.  The beam transmitted through the wedged beamsplitter is focused to form a beam waist of approximately 2~mm at the surface of the test mass.  It is then divided into two beams of equal power, with the beamsplitting ratio tuned by adjusting the angle of incidence on the beamsplitter.  The output beams enter a separate section of the transmitter housing that is designed to accommodate the {\em Working Standard} power sensor used for laser power calibration (see Sec.~\ref{Calibration}) and left-hand or right-handed  configurations for operation on either arm of the interferometer (see Fig ~\ref{fig:Txmodule_layout}).

The receiver module is shown schematically in Fig.~\ref{fig:Txmodule_layout}~(b). The Pcal beams reflected from the test mass and redirected by the in-vacuum periscope structure enter the receiver module and are directed by a pair of mirrors to a power sensor. This sensor is a 4 in.\ diameter integrating sphere with an InGaAs photodetector that collects both Pcal beams after reflection from the test mass and transmission through the output window.

The ratio of the power measured at the receiver module to that measured at the transmitter module gives the overall optical efficiency.  It is typically about 98.5\,\%.~\cite{T1500131} Using this optical efficiency, the power measured with either the transmitter or receiver photodiodes can be used to estimate the amount of laser power driving the test mass. Sec.~\ref{Calibration} describes the absolute calibration process for these power sensors.
\subsection{Optical Follower Servo}\label{IIB}
\begin{figure}
\includegraphics[width = .48\textwidth]{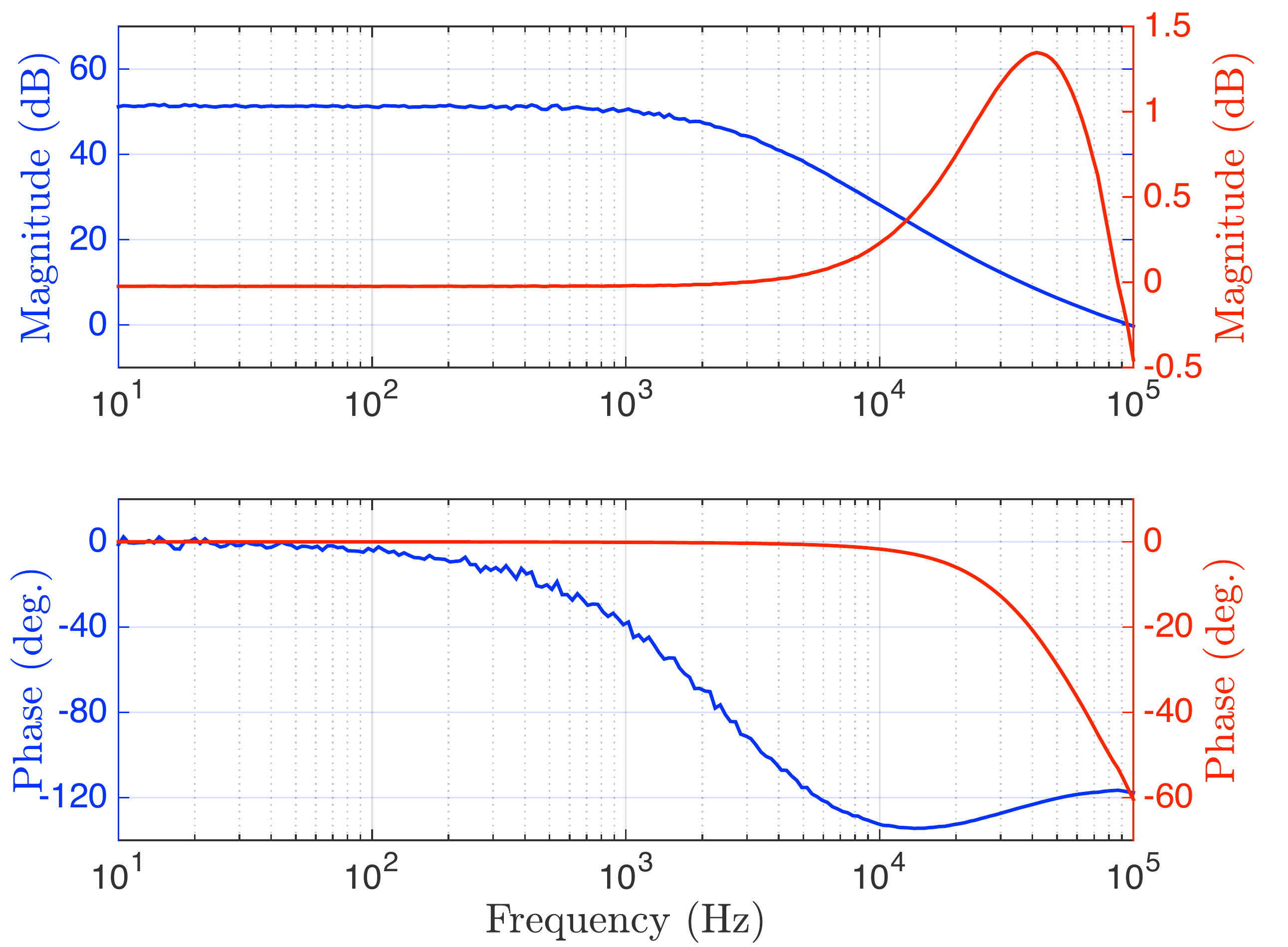}
\caption{Measured open-loop (blue) and closed-loop (red) transfer functions of the Optical Follower Servo. The unity gain frequency is approximately 100~kHz and the phase margin is about 62~deg.}
\label{fig:OFScltf}
\end{figure}

The open and closed loop transfer functions of the Pcal Optical Follower servo are shown in Fig.~\ref{fig:OFScltf}.
The unity gain frequency is approximately 100 kHz, with 62~deg. of phase margin.  At 5~kHz, the discrepancy between the requested and delivered sinusoidal waveforms is less than 0.005~dB (0.06\,\%) and the phase lag is approximately 0.6~deg.

\begin{figure}
\includegraphics[width = .5\textwidth]{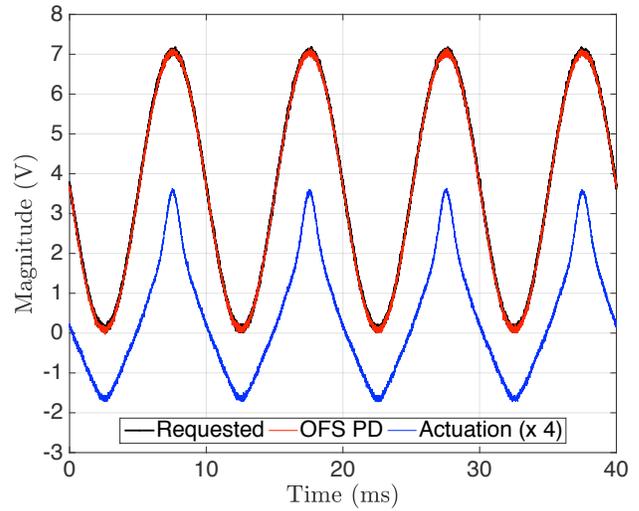}
\caption{Optical Follower Servo signals with the loop closed and modulating at 95\,\% of the maximum diffracted laser power.  The black trace (under the red trace) is the requested waveform.  The red trace is the delivered waveform measured by the OFS photodetector. The blue trace is the actuation signal (x 4) sent to the AOM driver.}
\label{fig:OFSDRIVE}
\end{figure}

\begin{figure}
\includegraphics[width = .5\textwidth]{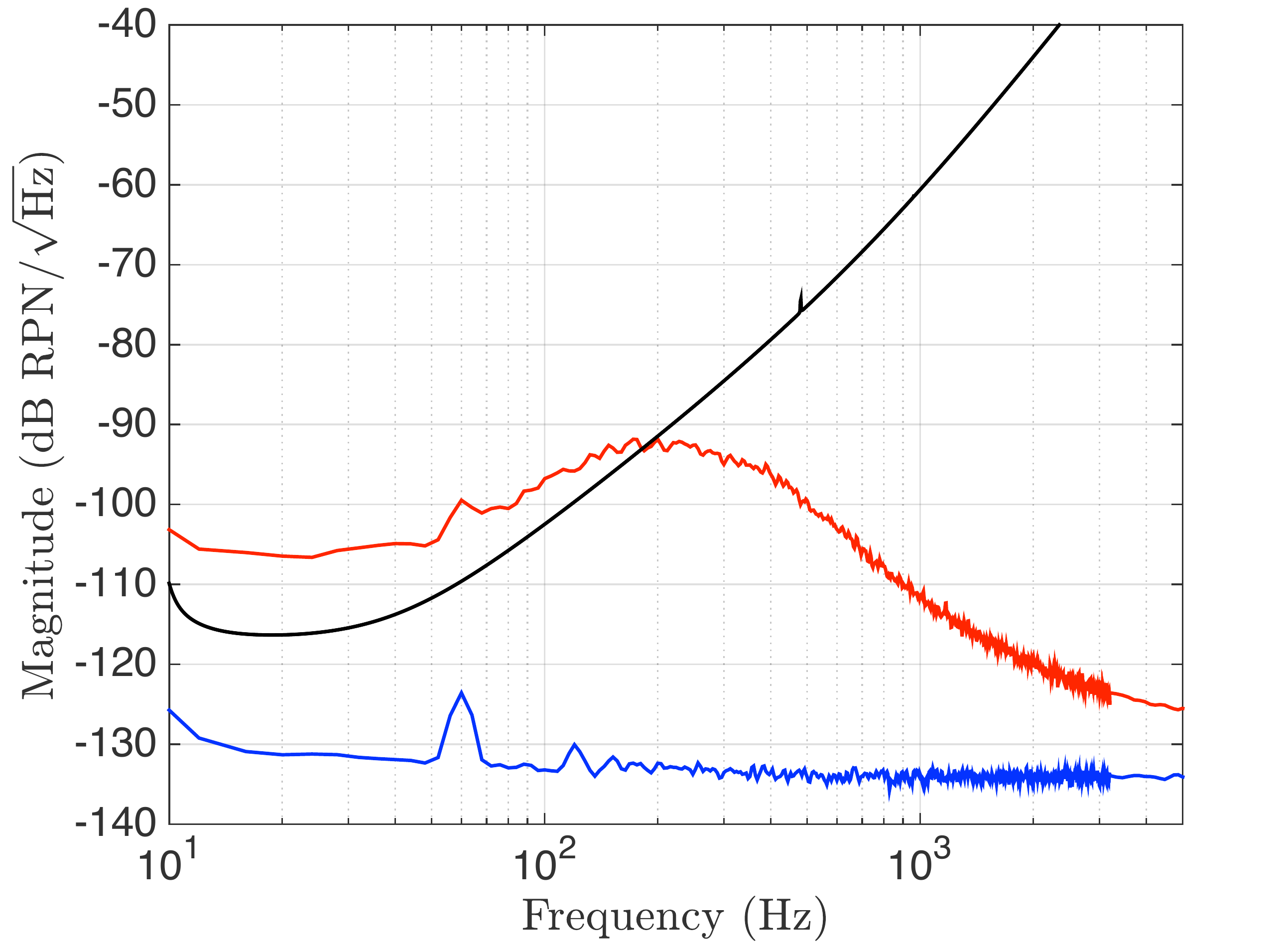}
\caption{Free running Relative Power Noise (RPN) of the Pcal laser (red) and the OFS suppressed RPN (blue). The suppressed RPN meets Advanced LIGO requirements (black) at frequencies above 10 Hz.}
\label{fig:OFS_RPN}
\end{figure}

\begin{figure}
\includegraphics[width = .5\textwidth]{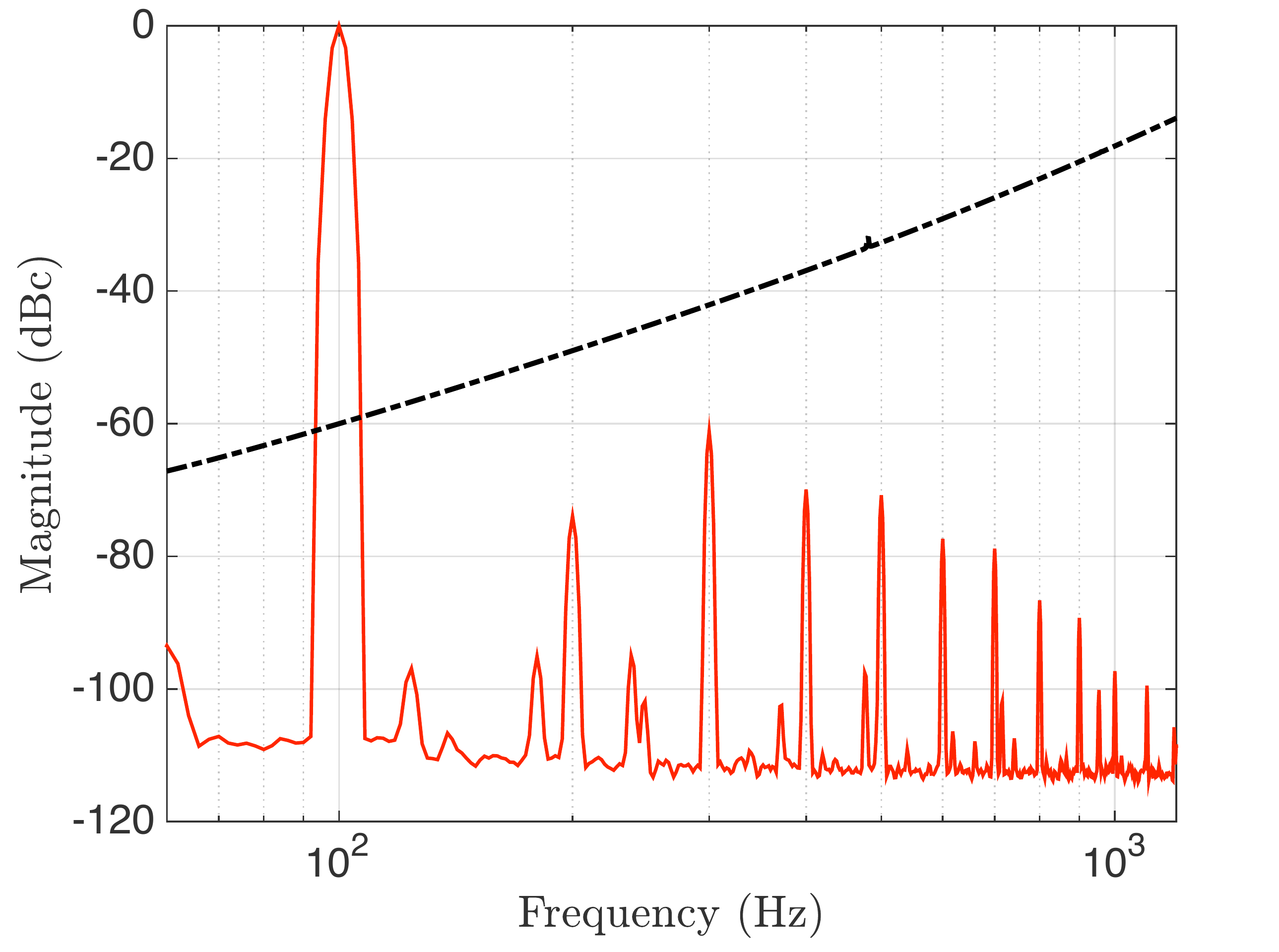}
\caption{Suppressed modulation harmonics relative to the carrier. The 100~Hz modulation is at 95\,\% of the maximum diffracted power.  All harmonics are well below the Advanced LIGO noise requirements (in black).}
\label{fig:harmSupp}
\end{figure}

This servo actuates the diffracted light level to ensure that the output of the OFS photodetector (see Fig.~\ref{fig:Txmodule_layout}) matches the requested modulation waveform.  It thus suppresses inherent laser power noise (see Fig.~\ref{fig:OFS_RPN}) as well as harmonics (see Fig.~\ref{fig:harmSupp}) of the requested periodic modulations that result from nonlinearity in the acousto-optic modulation process.  It enables operating with larger modulation depth without compromising performance, increasing actuation range by more effectively utilizing the available laser power.  Fig.~\ref{fig:OFSDRIVE} shows the waveform measured by the OFS photodetector (red trace) with the servo loop operating and modulating the maximum diffracted laser power by 96\,\% peak-to-peak.
The black trace (under the red trace) is the requested waveform and the blue trace is the actuation signal, multiplied by a factor of 4 for better visualization, sent to the AOM driver.

Fig.~\ref{fig:OFS_RPN} shows the free-running (in red) and OFS-suppressed (in blue) relative power noise (RPN) of the Pcal laser light. The suppressed power noise is well below the Advanced LIGO noise requirements at all frequencies that are of interest to LIGO.
Fig.~\ref{fig:harmSupp} shows the suppression of modulation harmonics relative to the carrier as detected by the outside-the-loop transmitted light power sensor for a requested sinusoidal waveform at 100 Hz and 95\,\% of the maximum modulation depth. The harmonics are well below the Advanced LIGO requirement, plotted in black. Furthermore, the modulated power required to achieve an SNR of 100 at 100~Hz is a factor of about 20 less than the maximum modulation and the sideband amplitudes are much lower for lower modulation amplitudes.

\begin{figure}
\includegraphics[width = .5\textwidth]{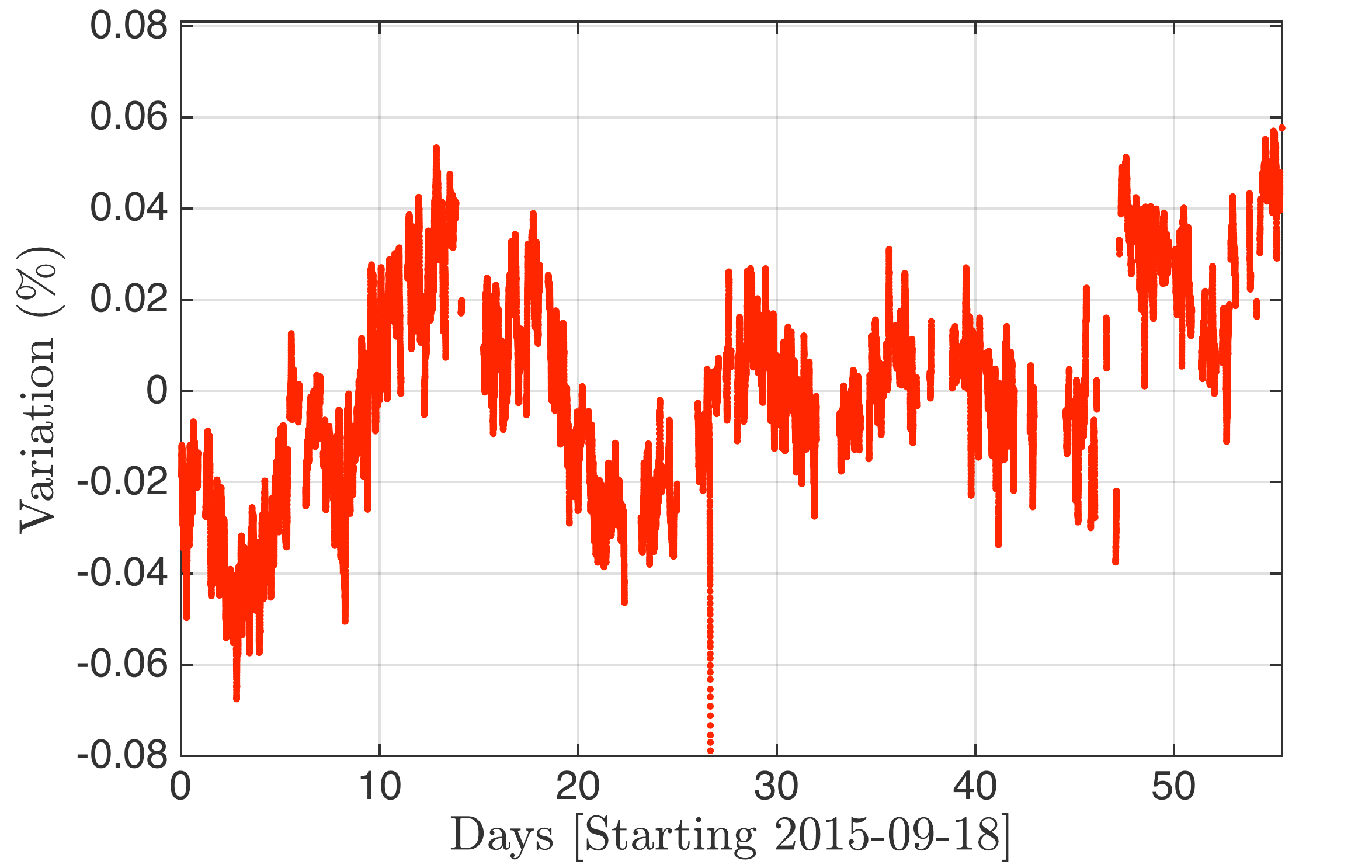}
\caption{Trend of the normalized amplitude of the power modulation measured by the power sensor in the receiver module. The amplitudes are calculated using Fourier transform with 60\,s integration interval.}
\label{fig:trendcurve}
\end{figure}

By injecting a constant amplitude waveform into the optical follower servo, the long term stability of the Pcal system can be evaluated by measuring the amplitude of the laser power modulation measured with the power sensor in the receiver module.  The amplitude of this signal measured over a sixty day interval is plotted in Fig~\ref{fig:trendcurve}. The peak-to-peak variation is approximately 0.1\,\%.
Measuring the waveform of the power reflected from the test mass and using it to estimate the induced motion eliminates errors caused by discrepancies due to the induced waveform not exactly matching the requested waveform. Thus, the data in Fig.~\ref{fig:trendcurve} represent an upper limit of the temporal variations in the Pcal calibration.
\subsection{Beam Localization System}\label{IIC}
In 2009, responding to the predictions of Hild, et al.,~\cite{P070074} Goetz, et al.\ demonstrated~\cite{iLIGocal} that Pcal errors could be as large as 50\,\% due to local deformation of the test mass surface.  This led to dividing the Pcal laser into two beams and positioning them away from the center of the mirror surface.  Induced rotation of the mirror is minimized by maintaining the center of force for the Pcal beams as close as possible to the center of the mirror surface. The location of Pcal center of force, $\vec{a}$, depends on the beam positions and the ratio of powers in the individual Pcal beams. It is given by 
\begin{equation}
\vec{a} = \frac{\beta \vec{a_{1}}+\vec{a_{2}}}{\beta + 1}
\label{eq:pwrbal}
\end{equation}
where $\vec{a}_{1}$ and $\vec{a}_{2}$ are the displacement vectors of the two Pcal beams about the center of the mirror face and $\beta = P_{1}/P_{2}$ is the ratio of beam powers.~\cite{P080118}. Reducing calibration uncertainties introduced by unwanted rotation can also be minimized by maintaining the the position of the main interfometer beam close to the center of the optic.  Both displacements enter Eq.~\ref{eq:pcaldisp} via the dot product in the term in square brackets.

In 2009, Daveloza et al.\ published the results of finite element modeling that showed that bulk elastic deformation resulting from Pcal forces can compromise the calibration, especially at frequencies above 1~kHz.~\cite{P1100166}  Their results for the Advanced LIGO test masses indicated that if the Pcal and interferometer beams are at their optimal locations the induced calibration errors would be less than 1\,\% at frequencies below 4.3~kHz. However, for significant offsets of the Pcal beams from their ideal locations these errors would increase dramatically at frequencies above $\sim$1~kHz . If the Pcal beams were displaced by a few millimeters, the errors could be as large as 10\,\% at 5~kHz.

\begin{figure}
\includegraphics[width = 0.42\textwidth]{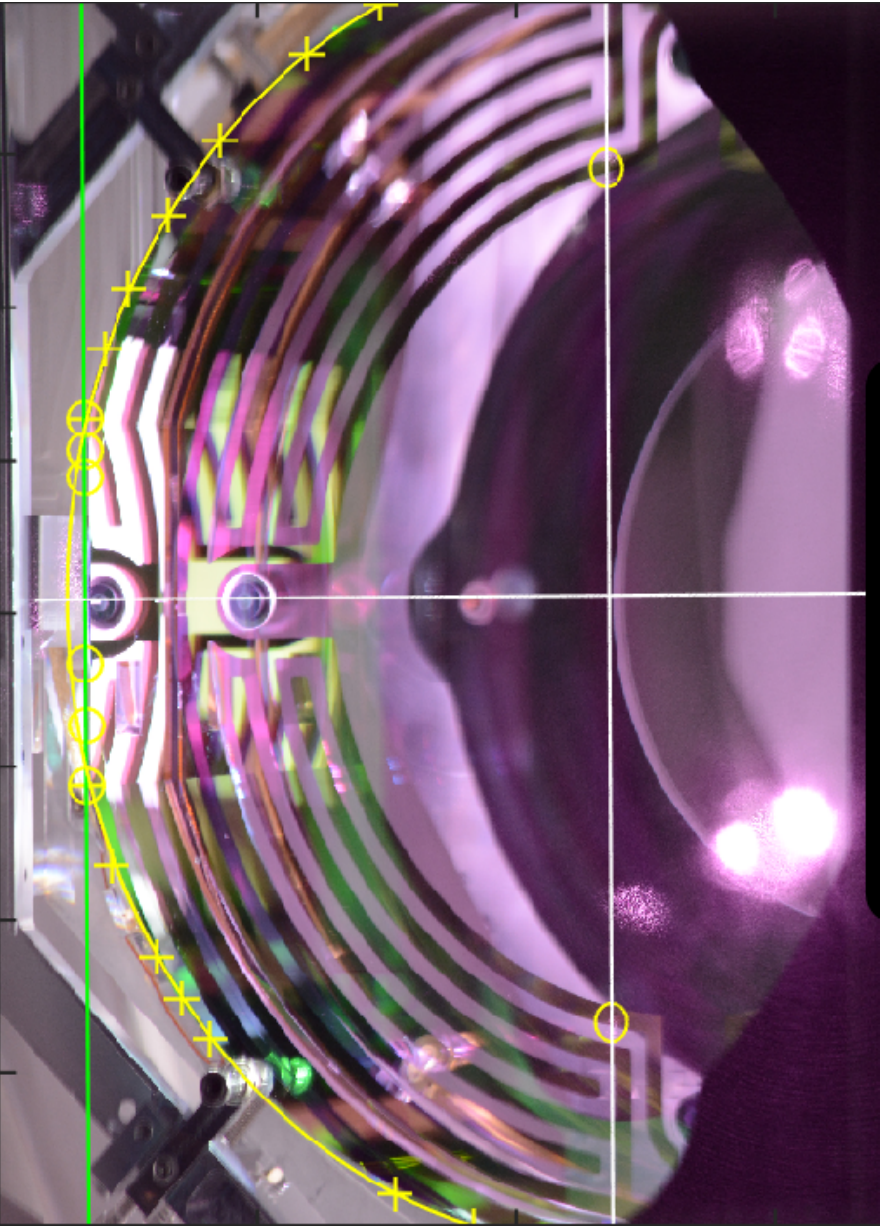}
\caption{Image of an end test mass from a Pcal beam localization camera system. The right side is occluded by the stray-light baffling. The mirrors have flats on the sides for attachment of the suspension fibers. These flats are oriented vertically and are used to determine the azimuthal orientation of the images. The well-defined angle of view along with the dimensions of the mirror enable determination of the beam positions on the mirror surface by identifying points on the edge of the optic (yellow crosses) and fitting the appropriate ellipse to the points. The system is designed to determine the optimal positions of the beams on the mirror surface (yellow circles above and below center) with millimeter accuracy.}
\label{fig:Pcal_beams}
\end{figure}

To determine the Pcal spot-positions, the Advanced LIGO Pcals use beam localization systems consisting of a high-resolution (6000 $\times$ 4000 pixels) digital, single lens reflex camera (Nikon D7100) with the internal infrared filter removed, a telephoto lens, and remotely controlled via an ethernet interface. The camera systems are mounted on separate vacuum ports, and use relay mirrors mounted to the same Pcal in-vacuum periscope structure to acquire images of the test mass surfaces such as the one shown in Fig.~\ref{fig:Pcal_beams}.
Points along the vertical flats on the sides of the mirror for attachment of the suspension fibers are used to orient the images azimuthally. Then, points along the edge of the mirror surface together with the well-defined angle of view and the dimensions of the mirror blank are used to fit the appropriate ellipse to the image and identify the coordinates of the center of the mirror (in pixel space). Pcal beam spot positions are determined by observing the scattered light from the Pcal beams in camera images. This information is used to direct the Pcal beams to their optimal locations, above and below the center of the optic, using the mirror mounts in the transmitter modules.
\section{Laser Power Sensor Calibration}\label{Calibration}
The absolute scale of the test mass displacement estimation, and therefore the overall interferometer response, is set fundamentally by the measurements of laser power in the transmitter and receiver module photodiodes. In this section we describe the propagation of absolute calibration from a single NIST-traceable Gold Standard to all eight photodiodes used thus far in Advanced LIGO (two per end-station, two end stations per interferometer, two interferometers). 
\subsection{Calibration Standards}
\begin{figure}[t]
\includegraphics[width = .5\textwidth]{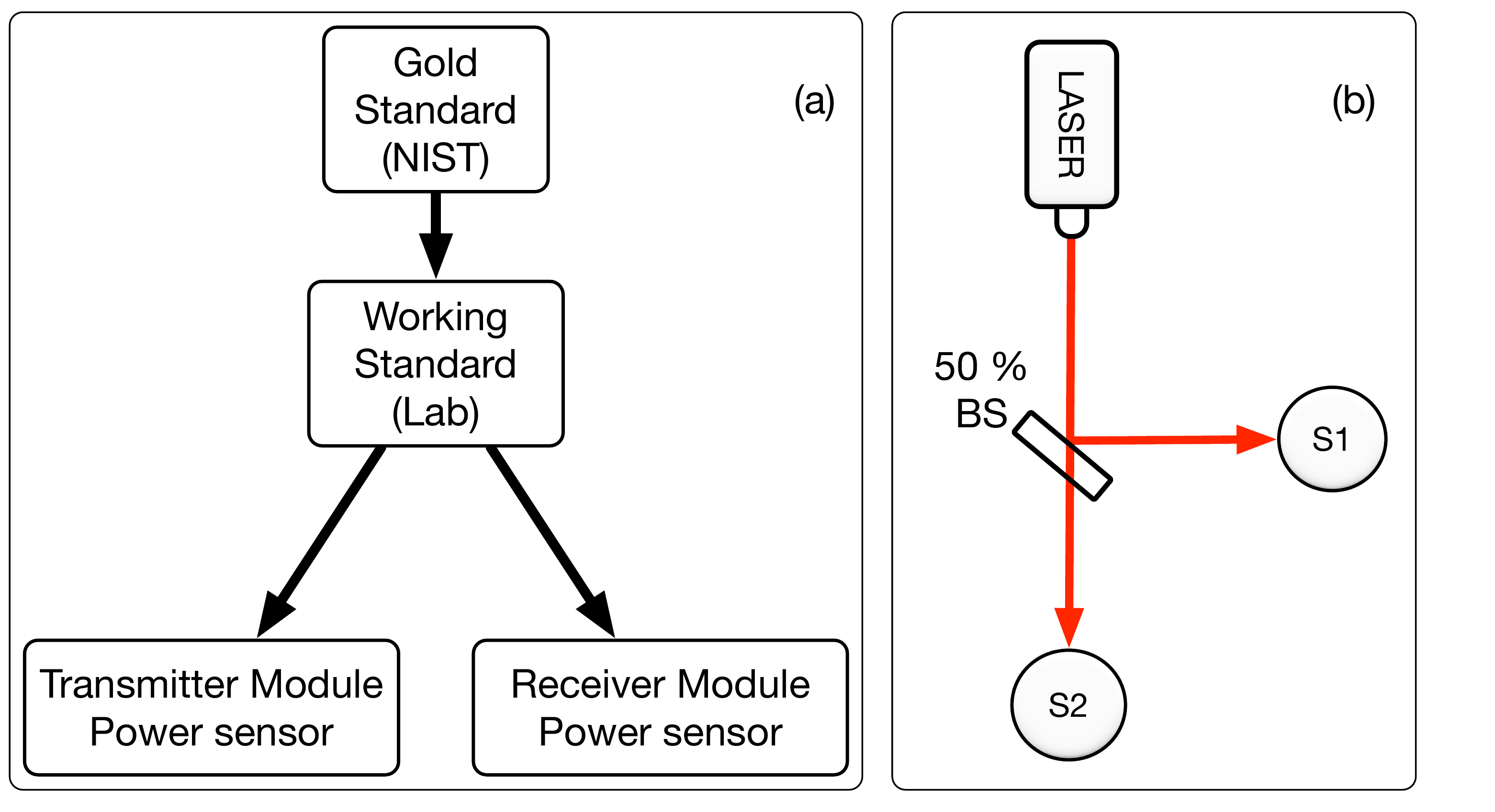}
\caption{(a) Schematic diagram of the chain of the calibration transfer from NIST to the Pcal laser power sensors.  (b) Schematic diagram of the setup used to transfer the calibration from the {\em Gold Standard} to a {\em Working Standard}. Each standard is placed alternately in the path of the reflected (R) and transmitted (T) beams to determine the ratio of the responsivities.}
\label{fig:caltransfer}
\end{figure}

Absolute laser power calibration is achieved using a power sensor referred to as the {\em Gold Standard} (GS) that is calibrated annually at the National Institute of Standards and Technology (NIST) in Boulder, CO.~\cite{T1500036} As shown schematically in Fig.~\ref{fig:caltransfer}a, the GS calibration is transferred to the power sensors in the  Pcal transmitter and receiver modules installed at the end stations via identical intermediary transfer standards, one per interferometer, referred to as {\em Working Standards} (WSs). The GS and WSs use unbiased InGaAs photodetectors mounted to  4~in.\ diameter integrating spheres.

\begin{figure}[t]
\includegraphics[width = .5\textwidth]{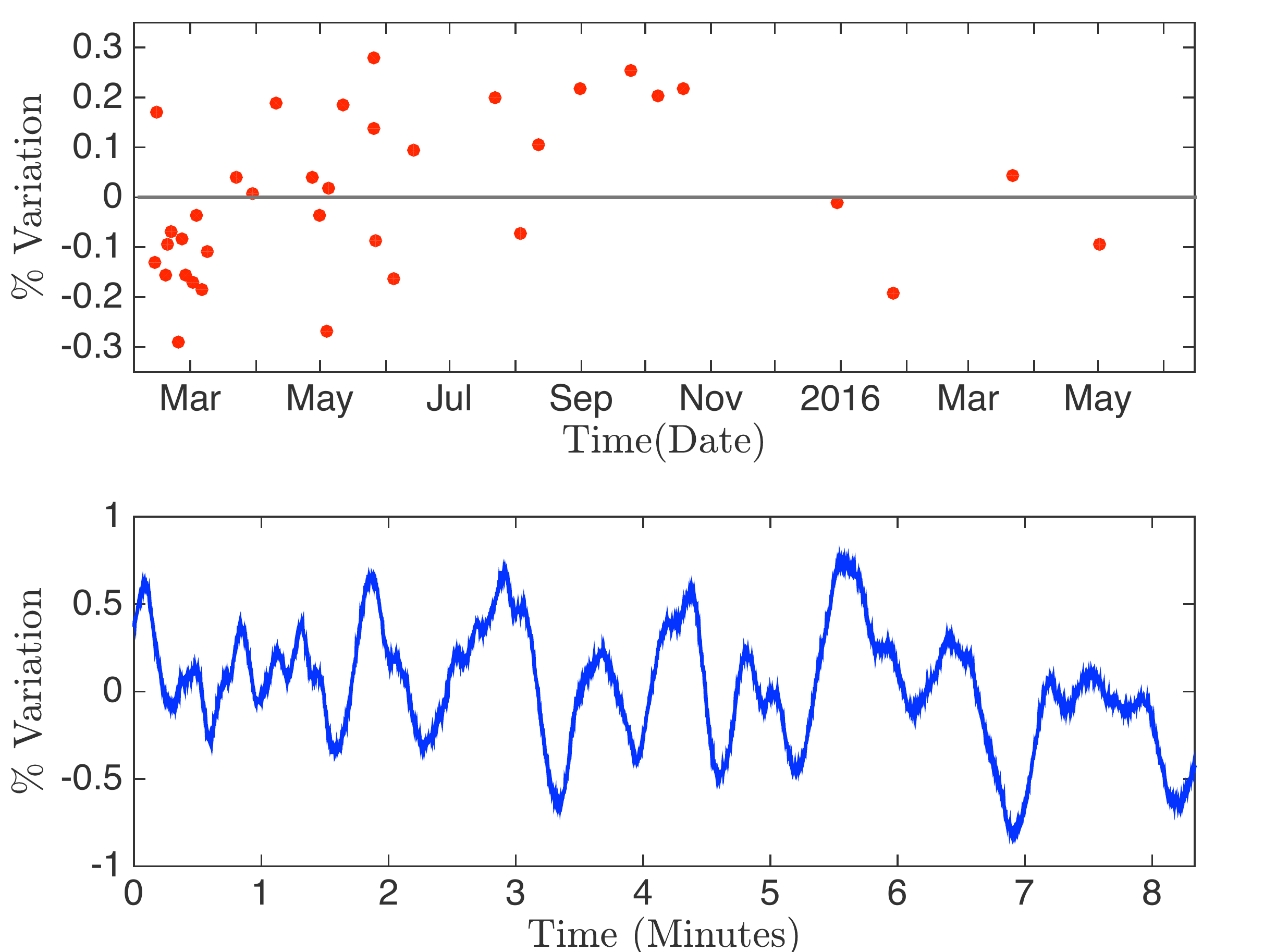}
\caption{Top: Working Standard over Gold Standard responsivity ratio measured over thirteen months. The maximum variation about the mean value is $\pm 0.3\,\%$; the standard deviation of the measurement is 0.14\,\% and the standard error of the mean from 36 measurements is 0.03\,\%. Bottom: A typical time series from one of the calibration standards showing the correlated output variations due to laser speckle.}
\label{fig:WS1_GS}
\end{figure}

The GS calibration is transferred to the WSs, using the experimental setup shown schematically in Fig.~\ref{fig:caltransfer}b.
The GS and a WS are alternately placed in the transmitted (T) and reflected (R) beams of the beamsplitter and time series of the detector outputs are recorded. The ratio between time series recorded simultaneously eliminates laser power variations and the ratio between the sets of time series eliminates the beamsplitter ratio, yielding the ratio of the  WS responsivity to that of the GS. These measurements are repeated periodically in order to track the long term stability of the standards.  The ratio of the Hanford WS to GS responsivities, measured over a thirteen month interval, is plotted in Fig.~\ref{fig:WS1_GS}~(top panel). During a typical measurement, slow variations in the signals of approximately 1\,\%  peak-peak with periods of tens of seconds are observed (see Fig.~\ref{fig:WS1_GS}, lower panel).  These are attributed to laser speckle in the integrating spheres.~\cite{T080316} Each measurement is recorded over a 10 minute interval and averaged in order to minimize the impact of laser speckle.
\subsection{End-Station Calibration}
The Working Standard (WS) at each observatory is used to calibrate the photodetectors inside the Pcal modules at each end station. The integrating sphere-based power sensors inside the transmitter and receiver modules are used to monitor the Pcal light power directed into and transmitted out of the vacuum envelope.  They thus place upper and lower bounds on the Pcal power reflecting from the end test mass, with the discrepancy attributed to optical losses in the vacuum windows, relay mirrors and the test mass itself.  In principle, these losses could be measured and quantified, but in practice access to the vacuum envelope to make the required measurements is extremely limited.  We thus use the mean of the incident and reflected power as an estimate of the power incident on the test mass and expand our uncertainty estimate to account for the finite optical efficiency (see Sec.~\ref{Uncertainties}).

Calibration of the Pcal power sensors proceeds by placing the WS in the path of one or both Pcal beams, either in the dedicated power measurement section of the transmitter module or by removing the receiver power sensor and replacing it with the WS, and recording time series of the power sensor signals.  The power measured by the two power sensors, as the power exiting the transmitter module ($P_{T}$) and the power collected at the receiver module($P_{R}$), are thus given by
\begin{subequations}
\begin{equation}
P_{T}  = \left(\frac{1}{\alpha_{T} \; \alpha_{W} \; \rho_{G}}\right) V_{T} 
\label{eq:e3a}
\end{equation}
\begin{equation}
P_{R}  = \left(\frac{1}{\alpha_{R} \; \alpha_{W} \; \rho_{G}}\right) V_{R}
\label{eq:e3b}
\end{equation}
\end{subequations}
where, $\alpha_{T}$ and $\alpha_{R}$ are the power sensors to WS responsivity ratios, $\alpha_{W}$ is the WS to GS responsivity ratio, $\rho_{G}$ is the GS responsivity (in V/W) measured at NIST, and $V_{T}$ and $V_{R}$ are the power sensor readings in volts.

The estimated power at the end test mass, $\mathcal{P}_{T}$ and $\mathcal{P}_{R}$, in terms of power measured by the transmitter module and receiver module power sensors are given by 
\begin{subequations}
\begin{equation}
\mathcal{P}_{T}  = \left(\frac{1+e}{2}\right) \;P_{T}
 \label{eq:e4a}
\end{equation}
\begin{equation}
\mathcal{P}_{R}	 = \left(\frac{1+e}{2e}\right) \; P_{R} 
   \label{eq:e4b}
\end{equation}
\end{subequations}
where $e$ = $P_{R}$/$P_{T}$ is the end station optical efficiency. The estimated power at the end test mass using either of the two power sensors gives the same result (i.e.\ $\mathcal{P}_{T} = \mathcal{P}_{R}$) and hence we will only use the power estimated by the receiver module power sensor, $\mathcal{P}_{R} = \mathcal{P} $, for uncertainty calculation in the Sec.~\ref{Uncertainties} below.

The photodetectors that are used for the Pcal power sensors were designed and fabricated by LIGO with particular attention given to maintaining a flat response over the band of frequencies from DC (NIST calibrations, and WS/GS responsivity measurements) up to 5~kHz.  They use InGaAs photodiodes operating in photovoltaic mode (unbiased).  Photocurrents are kept well below 1~mA.  To test the response of the receiver module power sensor, we temporarily installed a broadband commercial photodetector (NewFocus model M-2033) with an advertised bandwidth of over 200 kHz.  Driving the input to the OFS, we measured the ratio of the responses of the receiver module power sensor to that of the NewFocus photodetector.  Variations in the normalized ratio were less than $\pm$\,0.1\,\%  over the frequency range from 10~Hz to 5~kHz.~\cite{LLOalog}
\section{Uncertainties}\label{Uncertainties}
Several factors contribute to uncertainty in determining the displacements induced by the Pcals (see Eq.~\ref{eq:pcaldisp}). Laser power measurement is the most significant contributor to the overall uncertainty budget. The absolute power calibration of the Gold Standard, $\rho_{G}$, performed by NIST, has a 1-$\sigma$ uncertainty of 0.44\,\% for each measurement.~\cite{T1500036}  Combining the two most recent NIST measurements relevant for the current configuration of the GS, the 1-$\sigma$ relative uncertainty is 0.51\,\%.~\cite{T1500036} The {1-$\sigma$} relative uncertainty in the measured ratio of the Hanford WS responsivity to that of the GS ($\alpha_{W}$), based on 36 measurements made over a 13 month period (see Fig.~\ref{fig:WS1_GS}), is $0.03\,\%$. 

The subsequent transfer of the WS calibration to the Pcal power sensors involves six ratio measurements made with the WS at the end station. From these we determine the power sensor responsivity ratios, corrected for the Pcal optical efficiency, to estimate the power incident on the test mass,  $\alpha_{T}' = \left[2/(1+e)\right] \alpha_{T}$ and $\alpha_{R}' = \left[2e/(1+e)\right] \alpha_{R}$. The 1-$\sigma $ relative uncertainty (statistical only) associated with these measured quantities are typically smaller than 0.05\,\%. However, as described in Sec.~\ref{Calibration}, to account for the optical loss between the transmitter module and the receiver module, the power at the test mass is estimated by averaging the powers measured at the transmitter (upper limit) and receiver modules (lower limit). The actual value of the power at the test mass lies between these upper and lower limits and thus the uncertainty associated with optical efficiency is treated as a rectangular distribution (a Type B uncertainty, see NIST-1297~\cite{NIST-1297}). The 1-$\sigma$ relative uncertainty associated with the optical loss, $\sigma_{e}/e$, is thus $(1-e)/(2 \sqrt{3})$. 

The overall relative uncertainty in the estimate of the power that impinges on the test mass, measured by the receiver module power sensor is given by
\begin{equation}
	\begin{split}
		\frac{\sigma_{\mathcal{P}}}{\mathcal{P}}  = & \Biggl\{\frac{1}{3}\left(\frac{1-e}{2}\right)^2 + \left(\frac{\sigma_{\alpha_{R}'}}{\alpha_{R}'}\right)^2  \\
        & + \left(\frac{\sigma_{\alpha_{W}}}{\alpha_{W}}\right)^2 + \left(\frac{\sigma_{ \rho_{G}}}{\rho_{G}}\right)^2 \Biggr\}^\frac{1}{2}.
	\end{split}
\end{equation}
The components of this uncertainty estimate are summarized in Table~\ref{table:powerunc}.
\begin{table}
\renewcommand{\arraystretch}{1.4}
 \begin{tabular}{>{\raggedright}m{5cm} | C{3cm} } 
 \hline
 \textbf{Parameter}  &  \textbf{Relative Uncertainty} \\
 \hline\hline
 NIST \verb+->+ GS [$\rho_{G}$] & 0.51\,\% \\ 
 WS/GS [$\alpha_{W}$] & 0.03\,\%\\ 
 Rx/WS [$\alpha_{R}'$] & 0.05\,\% \\
 Optical efficiency [e] & 0.37\,\% \\ 
 \hline
 \textbf{Laser Power ($\mathcal{P}$)} & \textbf{0.57\,\%}  \\ 
 \hline
 \end{tabular}
\caption{Uncertainty estimate for the receiver module power sensor calibration in terms of power reflected from the end test mass. The NIST calibration and the optical efficiency are the most significant contributors to the uncertainty budget.}
\label{table:powerunc}
\end{table}

Another source of uncertainty is the angle of incidence at which the Pcal beams impinge on the test mass.  The incidence angle $\theta$, determined from mechanical drawings and tolerances, is 8.75~deg. Maximum deviations of the angle are bounded by the size of the periscope optics (2~in.\ diameter) that relay the beams to the end test mass.  The 1-$\sigma$ (Type B) relative uncertainty in the cosine of this angle is 0.07\,\%.

For frequencies above the suspension resonances, the displacement induced by the Pcals is inversely proportional to the mass of the test mass. The masses were measured before installation at each observatory using digital scales. The calibrations of these scales were tested using two 20~kg NIST-traceable reference masses. The measured mass determines the force-to-displacement transfer function, $S(f)$ in Eq.~\ref{eq:pcaldisp}, of the quadruple pendulum system. The measured mass has an uncertainty of $\pm 20$~g, which contributes to about 0.005\,\%, 1-$\sigma$ relative uncertainty. 
\begin{table}
\renewcommand{\arraystretch}{1.4}
 \begin{tabular}{>{\raggedright}m{5cm} | C{3cm}} 
 \hline
 \textbf{Parameter}  &  \textbf{Relative Uncertainty} \\
\hline\hline
Laser Power [$\mathcal{P}$] & 0.57\,\% \\
Angle [$\mathrm{cos}\theta$]  & 0.07\,\%\\
Mass of test mass [$M$] &  0.005\,\% \\  
Rotation [$(\vec{a}\cdot\vec{b})M/I$]  & 0.40\,\%\\ 
 \hline
 \textbf{Overall} & \textbf{0.75\,\%} \\ 
 \hline
 \end{tabular}
\caption{Uncertainty in Pcal induced length modulation $x(f)$ in Eq.~\ref{eq:pcaldisp}. The power calibration and the rotational effect introduce the most significant uncertainty. The rotational effect can be minimized by precise location of the Pcal beams.}
\label{table:dispunc}
\end{table}

\begin{figure}
\includegraphics[width = .35\textwidth]{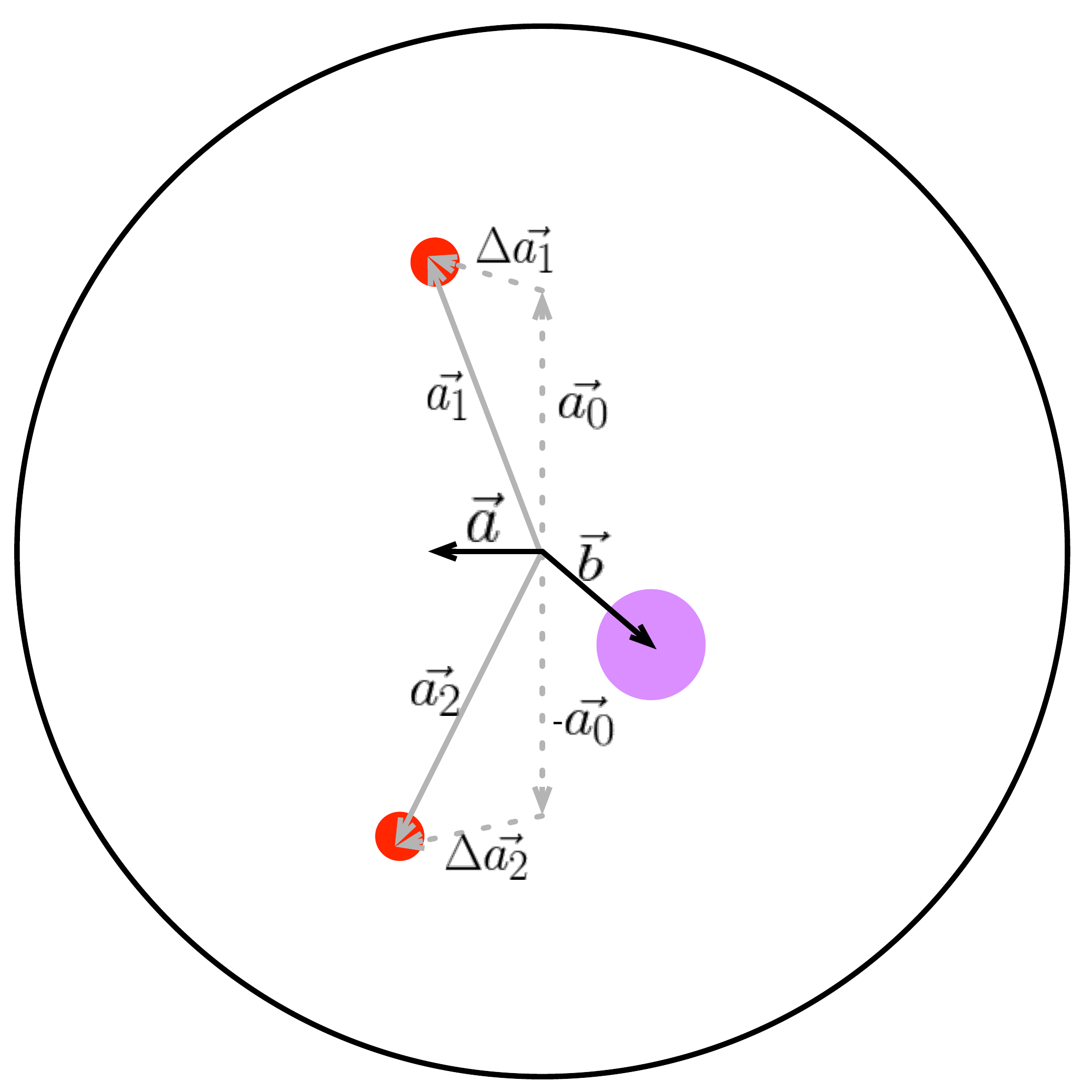}
\caption{Schematic showing the position of the Pcal and interferometer beams on the surface of the test mass. $\vec{a}$ and $\vec{b}$ are Pcal center of force and interferometer beam spot displacements from the center of the mirror surface. The beam positions and beam sizes are exaggerated for better illustration}
\label{fig:beamposition}
\end{figure}

A potentially significant source of uncertainty is apparent length changes sensed by the interferometer due to mirror rotation caused by offsets in the location of the interferometer and Pcal beams from their optimal positions. As described in Sec.~\ref{IIC}, the Pcal center of force depends on Pcal beam positions and power imbalance between the beams. Using $\vec{a_{1}} = \vec{a_{0}} + \Delta \vec{a_{1}}$ and $\vec{a_{2}} = -\vec{a_{0}} + \Delta \vec{a_{2}}$ as shown in Fig.~\ref{fig:beamposition} 
where $\lvert \vec{a_{0}}\rvert = 111.6$~mm is the magnitude of the nominal Pcal beam displacement from the center of the test mass and assuming that the effect of power imbalance on the beam offsets ($\Delta \vec{a_{1}}$ and $\Delta \vec{a_{2}}$) is minimal, we can write Eq.~\ref{eq:pwrbal} as:
\begin{equation}
\vec{a} \approx \vec{a_{0}}\left(\frac{\beta-1}{\beta+1}\right) +\left(\frac{\Delta \vec{a_{1}} + \Delta \vec{a_{2}}}{2}\right).
\label{eq:pcalforce}
\end{equation}
Using the position of the Pcal center of force, $\vec{a}$, calculated using Eq.~\ref{eq:pcalforce} above and the interferometer beam position $\vec{b}$, we can calculate the upper and lower limits of the uncertainty associated with the rotation effect, given by $\pm (\lvert\vec{a}\rvert\lvert\vec{b}\rvert) M/I$. Treating this as Type~B uncertainty, the 1-$\sigma$ uncertainty can be obtained by dividing the range defined by these limits by $2\sqrt[]{3}$.

Preliminary measurements indicate that the interferometer beam position offsets could be as large as $\pm$13~mm.~\cite{G1501362}.  The Pcal beam positions have been estimated using the Pcal beam localization systems described in Sec.~\ref{Hardware}. However, these estimates, which require identifying the center of the mirror surface in images that have poor contrast at the edge of the face of the optic, have not yet been optimized. Efforts to utilize the electrostatic actuator electrode pattern on the surface of the reaction mass that is positioned close to and behind the end test mass (see Fig.~\ref{fig:Pcal_beams}), rather than trying to identify the edge of the face of the test mass, are underway. A rough estimate of the maximum offset in the positions of the Pcal beams is $\pm$8~mm. Additionally, power imbalance also contributes to test mass rotation (see Eq.~\ref{eq:pcalforce}). The maximum measured power imbalance between the two beams is 2\,\%.

Using these estimates of interferometer and Pcal beam offsets, the maximum relative uncertainty introduced by rotation effects (see Eq.~\ref{eq:pcaldisp})is $\pm 0.70\,\%$.  Treating this as a Type~B uncertainty, the estimated 1-$\sigma$ relative uncertainty due to rotation effects is 0.40\,\%. This uncertainty can be reduced by positioning the Pcal beams more accurately.

Assuming negligible covariance between the components of the statistical uncertainty estimate, we combine the factors described above and listed in Table~\ref{table:dispunc} in quadrature. The estimated overall 1-$\sigma$ relative uncertainty in the Pcal-induced displacement of the test mass is 0.75\,\%.

A potential source of significant systematic uncertainty, especially at frequencies above $\sim$2~kHz, is the bulk elastic deformation described in Sec~\ref{Hardware}. Uncertainty due to this effect is not included in the analysis presented here. However it is being investigated and will be reported in future publications.
\section{Application}\label{Applications}
During normal interferometer operations, the Pcal systems at the ends of both arms operate continuously, injecting Pcal excitations at discrete frequencies, to support the calibration of the interferometer output signals. They are also periodically used to measure detector parameters -- sensing function, actuation function, signs and time delay -- that impact the calibrated output signals. These measurements are used to improve the calibration accuracy. Details of the Photon Calibrator measurements and operation are described below.

\subsection{Calibration Lines}
The excitations induced using the Pcals are also referred to as {\em{Calibration Lines}}. The nominal frequencies and amplitudes of these Pcal excitations are listed in Table~\ref{table:LineTable}. 
\begin{table}[!ht]
\begin{tabular}{C{1.5cm} | C{1.5cm} | C{2.25cm}| C{2.25cm} }
\hline
\multirow{2}{1.5cm}[-1.5ex]{\centering \textbf{Freq. (Hz)} } & \multirow{2}{1.5cm}[-1.5ex]{\centering \textbf{DFT Length (sec)} } & \multicolumn{2}{c}{ \textbf{Required Pcal Power} } \\ \cline{3-4}
 & &\textbf{Sept. 2015 Sensitivity} &\textbf{Design Sensitivity}\\
\hline
\hline
36.7  & 10 & 0.3\,\% & 0.1\,\%\\
331.9  & 10 & 10\,\% & 4\,\%\\
1083.7 & 60 & 77\,\% & 24\,\%\\
3001.3 & 3600 & 200\,\% & 50\,\%\\
 \hline
\end{tabular}
\caption{Photon Calibrator excitation frequencies during normal interferometer operations in Sept.~2015.  DFT intervals and  percentage of available laser power required to generate the excitations with SNR of 100, for the Sept. 2015 sensitivity and the Advanced LIGO design sensitivity.}
\label{table:LineTable}
\end{table}
The two lowest frequency excitations, near 37 and 332~Hz, are used in both the output signal calibration process and for tracking slow temporal variations. Applying corrections for these slow temporal variations improves calibration accuracy.~\cite{T1500377}  The SNR of approximately 100 is required to enable calibration at the one percent level with 10-second integration intervals.  The excitations near 1.1~kHz and 3~kHz are used to investigate the accuracy of the calibration at higher frequencies using longer integration times.  The excitation frequencies were chosen to avoid known potential sources of gravitational wave signals (rapidly-rotating neutron stars observed electromagnetically as pulsars), and to most effectively determine key interferometer parameters while avoiding the most sensitive region of the detection band.

Table~\ref{table:LineTable} also lists the percentage of available Pcal modulated laser power required to achieve an SNR of 100 with the listed discrete Fourier transform (DFT) time for each excitation.  The three lowest frequency lines are generated using the Pcal system at one end station.  The 3~kHz line is generated using the Pcal system at the other end station and consumes more than half of the available modulated power to achieve an SNR of 100 with DFTs of one hour at design sensitivity.  DFTs of more than 4 hours duration were required to reach this SNR with the Sept.~2015 sensitivity.

The amplitude of laser power modulation required to induce a length modulation with a desired SNR is given by
\begin{equation}
P (f_{i}) = \frac{c}{2 \cos \theta \, S(f_{i})}\frac{\Delta L(f_{i})\;\textrm{SNR}(f_{i})}{\sqrt{T}}
\end{equation}
where $f_{i}$ is the modulation frequency, $\Delta L(f_{i})$ is the amplitude spectral density of the interferometer sensitivity noise floor, and $T$ is the measurement integration time.

\begin{figure}[t]
\includegraphics[width = .5\textwidth]{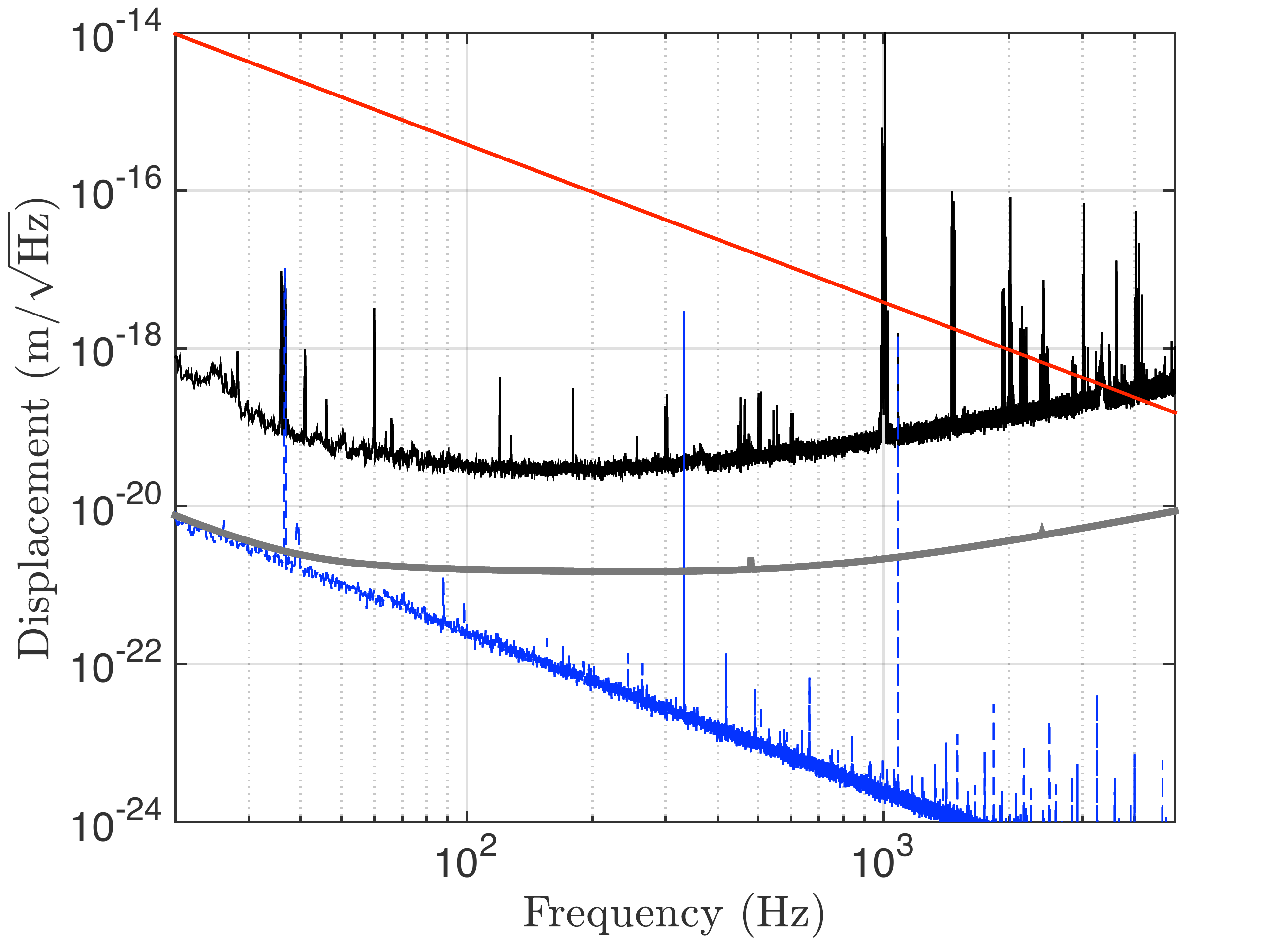}
\caption{Maximum modulated displacement using all of the available Photon Calibrator power at one frequency (red).  Pcal-induced displacements in Sept.~2015 (blue) along with the Sept.~2015 sensitivity noise floor (black) with a 10 second integration time. The gray curve is the maximum allowed unintended displacement noise, one tenth of the design sensitivity noise floor.}
\label{fig:pcalnoisecurve}
\end{figure}

For the Advanced LIGO Pcals the amplitude spectral density of the maximum modulated displacement that can be achieved using all of the available Pcal laser power is plotted in Fig.~\ref{fig:pcalnoisecurve} for a 10-second integration interval.
It falls as $1/f^{2}$ due to the force-to-displacement response from $1 \times 10^{-14}$~m/$\sqrt{\textrm{Hz}}$ at 20~Hz to below $2 \times 10^{-19}$~m/$\sqrt{\textrm{Hz}}$ at 5~kHz.  Fig.~\ref{fig:pcalnoisecurve} also shows the displacements induced by the Pcal excitation and the interferometer noise floor.  Finally, the requirement for the maximum unwanted Pcal-induced displacement noise, one tenth of the design sensitivity noise floor, is plotted.  As the interferometer sensitivity improves and the noise floor approaches the design levels, the amplitude of the Pcal excitations can be reduced proportionately, reducing the laser power required and therefore also the level of unwanted displacement noise. 

\begin{figure}[t]
\includegraphics[width = .5\textwidth]{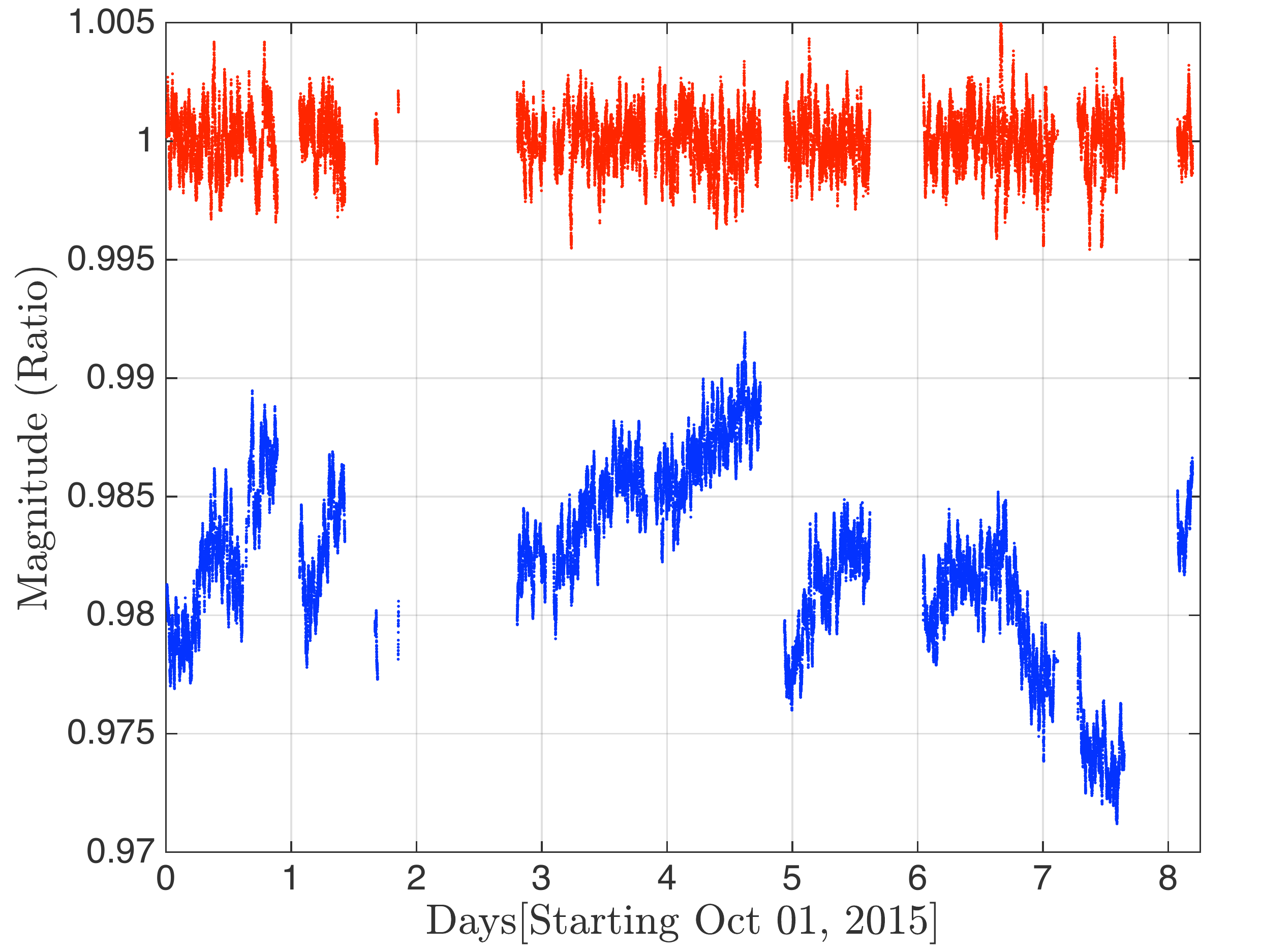}
\caption{Trends of the ratio between the displacement from the calibrated interferometer output signal and the calculated displacement from the Pcal power sensor in the receiver module using the excitation at 332~Hz. Blue: uncorrected data showing the slow temporal variations in the interferometer parameters. Red: corrected data after applying the calculated time-varying correction factors.}
\label{fig:gdspcal1khz}
\end{figure}

Pcal excitations are also used to monitor slow temporal variations in the response of the interferometers to differential length variations.  The frequencies of the excitations were selected in order to optimize this capability.  The slow variations in the interferometer calibration, measured using a Pcal line near 332~Hz,  over an eight day period in Sept.~2015 are shown in Fig.~\ref{fig:gdspcal1khz}.  The slow variations in the calibrated output signal are as large as 3\,\%.  Also shown in Fig.~\ref{fig:gdspcal1khz} are the calibration data that were corrected for the observed slow variations using calibration parameters calculated using the Pcal excitations.~\cite{T1500377}  On-line calculation and compensation for the time varying parameters using the Pcal lines is being implemented for future LIGO observing campaigns.  

\subsection{Frequency Response Measurements}
\begin{figure}[t]
\includegraphics[width = .5\textwidth]{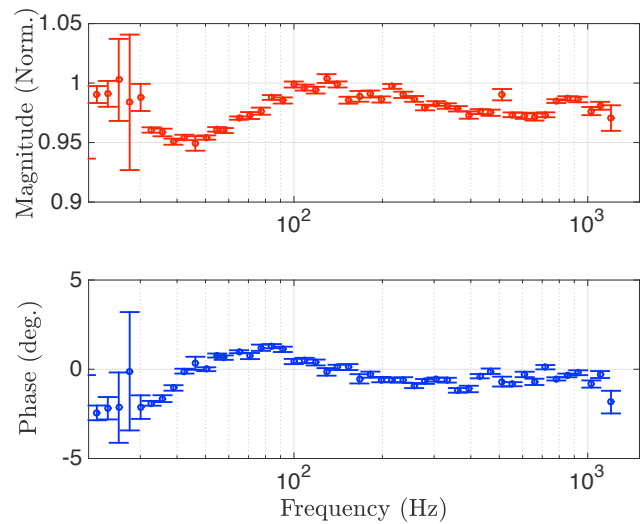}
\caption{Magnitude and phase of a typical swept-sine measurement of the transfer function between displacement induced (and calibrated) by the Pcal and the calibrated output of the interferometer.}
\label{fig:pcal2darmsweep}
\end{figure}

To assess the accuracy of interferometer calibration over a wide range of frequencies, swept-sine measurements are made by varying the Pcal laser power modulation frequency and measuring the complex response of the calibrated interferometer output signals.  These measurements are made during dedicated calibration interludes, the length of which are minimized in order to maximize observing time.  Thus, the Pcal displacement amplitudes must be sufficiently large to complete the measurements in a relatively short time.  Fig.~\ref{fig:pcal2darmsweep} shows a typical transfer function from 20~Hz to 1.2~kHz, with approximately 60 points.
The measurement was made in approximately one hour; the measurement statistical uncertainties, calculated from the coherence of the measurements, are approximately 1\,\% in amplitude and 1~deg.\ in phase, for frequencies between 30~Hz to 1.2~kHz. The statistical variation are higher in the band from 20-30~Hz due to resonances in the suspension systems of ancillary interferometer optics.   

\begin{figure}[!ht]
\includegraphics[width = .5\textwidth]{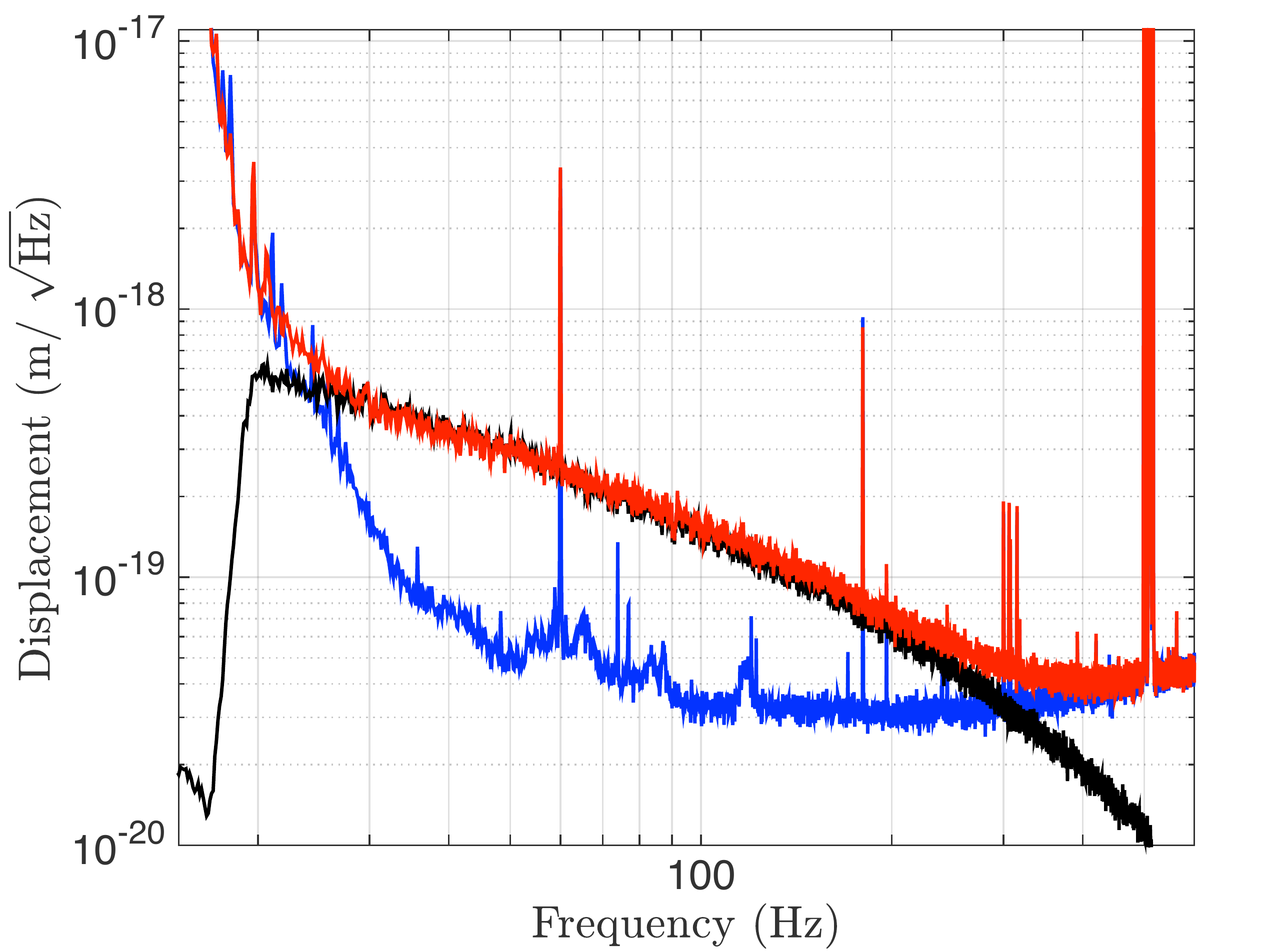}
\caption{Pcal broadband displacement excitation (black) and calibrated interferometer output signal both with (red) and without (blue) the Pcal excitation.}
\label{fig:pcalbroadband}
\end{figure}

Rather than injecting Pcal excitations at discrete frequencies, the transfer function can also be measured simultaneously by injecting a broadband signal. This can potentially make the calibration comparison process faster and more accurate.  It also has the potential of revealing features in the transfer function that might be missed in measurements made at only discrete frequencies. However, this type of measurement is also limited by the available Pcal laser power. To assess the feasibility of this method, a broadband signal covering the 30-300~Hz frequency band, band-pass filtered to attenuate it at higher and lower frequencies, was injected into the Pcal Optical Follower Servo. Fig.~\ref{fig:pcalbroadband} shows the displacement injected by the Pcal together with the calibrated interferometer output signal both with and without the Pcal excitation.  
As the sensitivity of the interferometers improves, the band over which this method is useful will increase. No unexpected discrepancies, that might have been missed by the discrete-frequency transfer function measurement were identified.


\subsection{Differential-mode and Common-mode Actuation}
Normally, the differential length response of the detector is calibrated using one Pcal system, varying the length of only one arm.  The Advanced LIGO interferometers, however, have Pcal systems installed at both end stations. They can be used simultaneously to produce either pure differential arm length variations, where the two arms of the interferometer stretch and contract out of phase or pure common arm-length variations, where the arms stretch and contract in phase.  Comparing differential and common excitations, enables diagnosing systematic differences between the two arms and quantifying the coupling between common-arm motion and differential-arm motion.

\begin{figure}
\includegraphics[width = .50\textwidth]{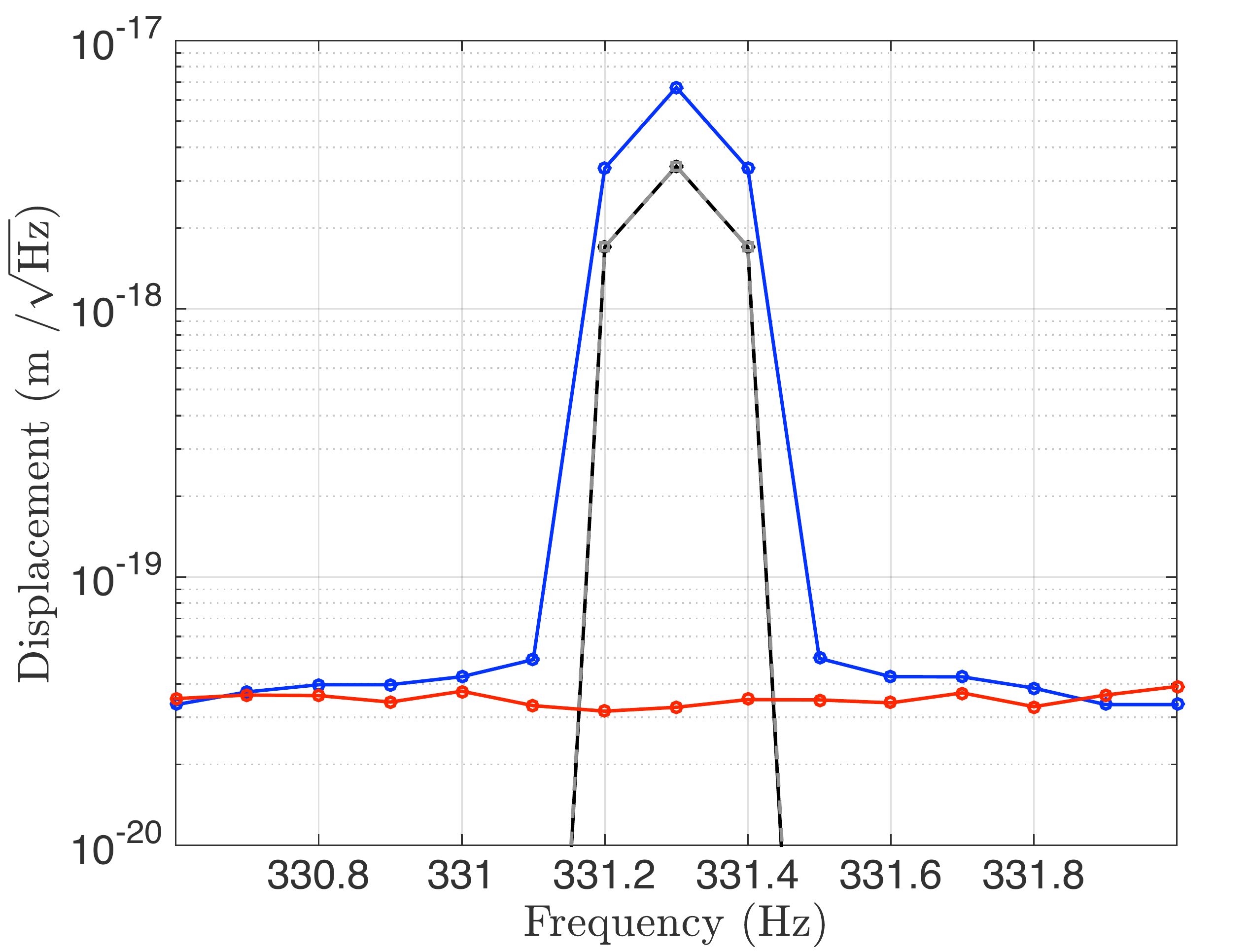}
\caption{Measurement using the Pcal modules at both end stations to induce equal-amplitude modulation of the positions of the test masses (overlapping gray and black) in {\em common mode} (red), 0~deg.\ relative phase, and {\em differential mode} (blue), 180~deg.\ relative phase.}
\label{fig:darmcarm}
\end{figure}

A comparison of differential and common actuation of the Livingston interferometer using the Pcals is shown in Fig.~\ref{fig:darmcarm}.
Both Pcal systems induced modulated displacements of equal amplitudes, as determined by the calibration of the Pcal receiver module power sensors.  The relative phases of the excitations was changed from 0~deg.~(in phase) to 180~deg.~(out of phase) to transition between common and differential actuation. Less than 0.2\,\% of the common-mode motion, within the measurement uncertainty, is sensed as differential mode motion by the interferometer. 

The ability to precisely vary the amplitude and phase of the injected length modulations enables high-precision calibration measurements without inducing large amplitude lines in the output signal. This can be realized by canceling length excitations injected by other actuators with Pcal lines injected at the same frequency but 180~deg.~out of phase. 

\subsection{Measuring Time Delays and Signs }
Radiation pressure actuation via the Pcals has a simple phase relationship between the length excitation (modulated laser power detected by the receiver module power sensor) and the induced motion of the test mass. For frequencies much larger than the 1~Hz resonances of the test mass suspension system, the induced motion of the test mass is 180 deg. out of phase with respect to the excitation signal. This property of Pcal excitations was exploited for the initial LIGO detectors to investigate the sign of the calibrated interferometer output signals.~\cite{timingiligo} Confirming the relative signs of the interferometer outputs is crucial for localizing the source of the detected gravitational waves on the sky using two or more detectors.
\begin{figure}
\includegraphics[width = .5\textwidth]{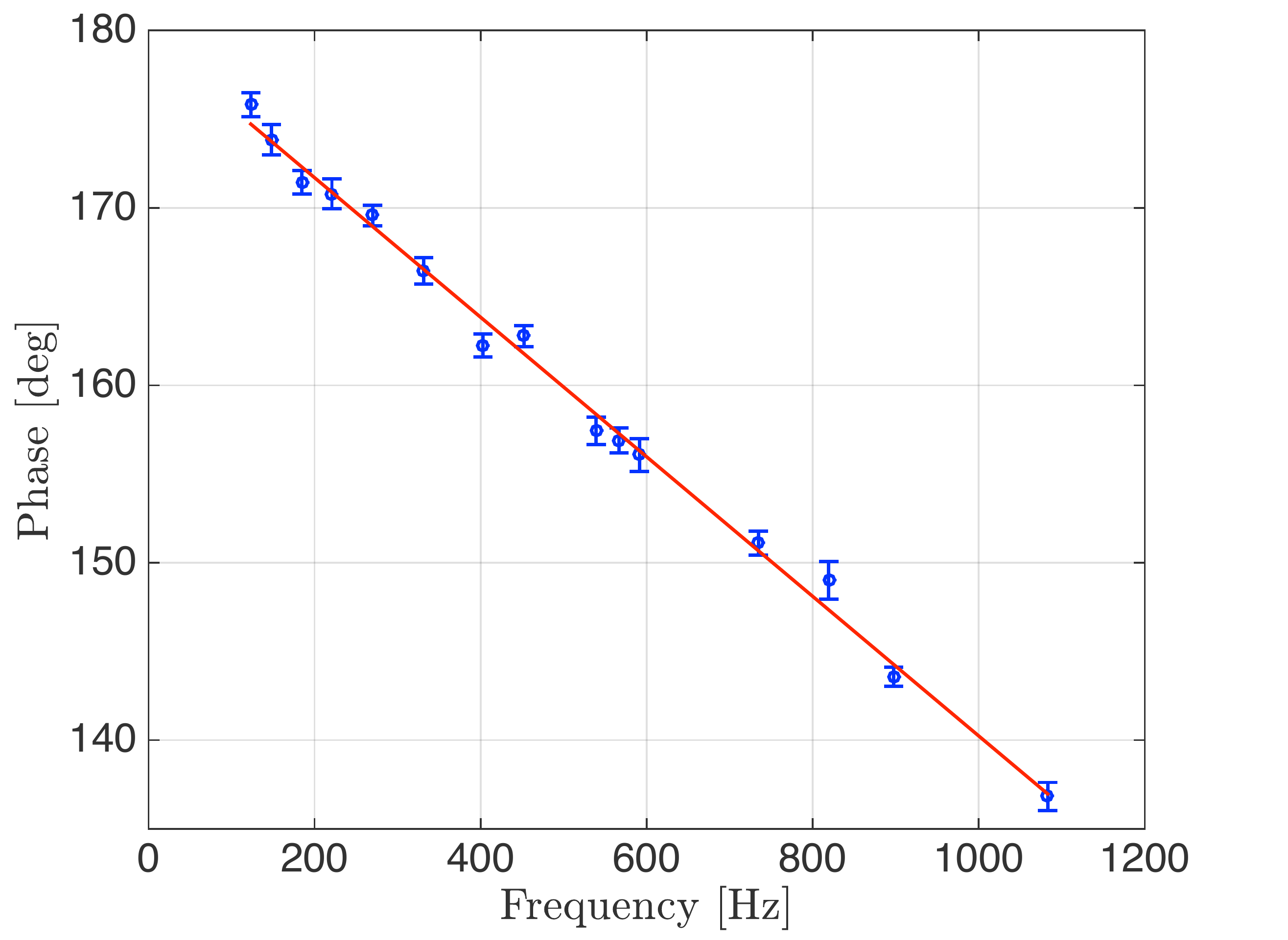}
\caption{Interferometer output signal timing measured using Pcal excitations. The least squares fit to the data shows the expected 180~deg.\ phase shift at low frequency and a delay of $109.2\pm2.2$~$\mu$s.}
\label{fig:timedelay}
\end{figure}

In addition to identifying the sign of signals, by using multiple excitations we can measure the time delays in the response of the detectors to motion of the test masses (and consequently gravitational waves). These delays also impact the sky localization of GW sources. Previously in LIGO, two frequencies were used to measure the delays yielding timing uncertainties on the order of 10~$\mu$s.~\cite{timingiligo} With the upgraded Advanced LIGO Pcal data acquisition and better timing standards, similar measurements are easily performed at many frequencies, or even broadband, and achieve measurement uncertainties of the order a few $\mu$s. Fig.~\ref{fig:timedelay} shows the results of signal delay measurements made at frequencies between 100 and 1100~Hz.
The straight line fit to the data shows the expected 180~deg. relative phase at lower frequencies and a time delay of $109.2\pm2.2$~$\mu$s. This delay arises from the combination of the effects of digital data acquisition ($76~\mu s$), analog electronics ($20~\mu s$) and light travel time in the arm ($13~\mu s$). The results of measurement like these are used to model the response of the interferometer to gravitational waves.~\cite{P1500248}
\section{Conclusions}\label{Conclusions}
The Advanced LIGO photon calibrators incorporate a number of upgrades that make them suitable for second generation gravitational wave detectors.  These include higher power lasers, low-loss vacuum windows, beam relay periscopes, optical follower servos, beam localization cameras, and receiver modules that capture the laser light reflected from the test masses.  One Pcal system is installed at each end station.  This enhances reliability by providing redundancy and provides additional actuation capabilities including increased range and the ability to make coordinated excitations.

The Pcal systems are now the primary calibration reference for the Advanced LIGO detectors, providing overall system uncertainty of 0.75\,\%. They are being used to track slow temporal variations in interferometer parameters that include optical gain, coupled-cavity pole frequency, and actuation strength. The resulting correction factors are being used to reduce errors in the calibrated interferometer output signals. 

Application of the Photon Calibrators is expanding to include injection of simulated gravitational wave signals in order to test the computer codes that search for signals in the LIGO data streams.~\cite{HWINJ}  Future uses may include actuation of the differential length degree of freedom to potentially reduce actuation drifts and noise and increase actuation range. \cite{pcaldarmactuation} As the Advanced LIGO sensitivity improves, and therefore the rate of detection of gravitational wave signals increases, better interferometer calibration accuracy and precision will be required in order to optimally extract source information from the signals. The photon calibrator systems are playing a key role in the ongoing efforts to reduce calibration uncertainties.
\section*{Acknowledgments}
LIGO was constructed by the California Institute of Technology and Massachusetts Institute of Technology with funding from the National Science Foundation, and operates under cooperative agreement PHY-0757058. Advanced LIGO was built under award PHY-0823459. Fellowship support from the LIGO Laboratory for both S.~Karki and D.~T.  and for D.~T. from the UTRGV College of Sciences is gratefully acknowledged. This paper carries LIGO Document Number LIGO-P1500249.
\bibliographystyle{unsrt}
\bibliography{Pcal}

\begin{thebibliography}{10}

\bibitem{detection}
B.~P. Abbott et~al.
\newblock Observation of gravitational waves from a binary black hole merger.
\newblock {\em Phys. Rev. Lett.}, 116:061102, Feb 2016.

\bibitem{detection2}
B.~P. Abbott et~al.
\newblock {GW}151226: Observation of gravitational waves from a 22-solar-mass
  binary black hole coalescence.
\newblock {\em Phys. Rev. Lett.}, 116:241103, June 2016.

\bibitem{P1600088}
B.~P. Abbott et~al.
\newblock Binary black hole mergers in the first {A}dvanced {LIGO} observing
  run.
\newblock {\em LIGO Document Control Center}, P1600088, 2016.

\bibitem{P1500248}
B.~P. Abbott et~al.
\newblock Calibration of the {A}dvanced {LIGO} detectors for the discovery of
  the binary black-hole merger {GW150914}.
\newblock {\em Phys. Rev. D}.

\bibitem{P1500218}
B.~P. Abbott et~al.
\newblock Properties of the binary black hole merger {GW150914}.
\newblock {\em LIGO Document Control Center}, T1500218, 2015.

\bibitem{Lindblom}
L.~Lindblom.
\newblock Optimal calibration accuracy for gravitational-wave detectors.
\newblock {\em Phys. Rev. D}, 80:042005, Aug 2009.

\bibitem{detectorpaper}
J.~Aasi et~al.
\newblock Advanced {LIGO}.
\newblock {\em Classical Quantum Gravity}, 32(7):074001, 2015.

\bibitem{mizuno}
G.~Heinzel et~al.
\newblock An experimental demonstration of resonant sideband extraction for
  laser-interferometric gravitational wave detectors.
\newblock {\em Physics Letters A}, 217:305--314, 1996.

\bibitem{P1500260}
D.~Martynov et~al.
\newblock The sensitivity of the {A}dvanced {LIGO} detectors at the beginning
  of gravitational wave astronomy.
\newblock {\em LIGO Document Control Center}, P1500260, 2016.

\bibitem{sei}
F.~Matichard et~al.
\newblock Seismic isolation of {A}dvanced {LIGO}: {R}eview of strategy,
  instrumentation and performance.
\newblock {\em Classical Quantum Gravity}, 32, 2015.

\bibitem{suspension}
S.M. Aston et~al.
\newblock Update on quadruple suspension design for advanced {LIGO}.
\newblock {\em Classical Quantum Gravity}, 29:305--314, 2012.

\bibitem{virgopcal}
T.~Accadia and {VIRGO} Collabration.
\newblock Reconstruction of the gravitational wave signal h(t) during the
  {VIRGO} science runs and independent validation with a photon calibrator.
\newblock {\em Class. Quantum Grav.}, 31:165013, 2014.

\bibitem{Glasgowpcal}
D.~A. Clubley et~al.
\newblock Calibration of the {G}lasgow 10 m prototype laser interferometric
  gravitational wave detector using photon pressure.
\newblock {\em Phys. Lett. A}, 283:85, 2001.

\bibitem{GEO600pcal}
K.~Mossavi et~al.
\newblock A photon pressure calibrator for the {GEO600} gravitational wave
  detector.
\newblock {\em Phys. Lett. A}, 353:1, 2006.

\bibitem{P080118}
E.~Goetz et~al.
\newblock Precise calibration of {LIGO} test mass actuators using photon
  radiation pressure.
\newblock {\em Class. Quantum Grav.}, 26:245011, 2009.

\bibitem{iLIGocal}
E.~Goetz et~al.
\newblock Accurate calibration of test mass displacement in the {LIGO}
  interferometers.
\newblock {\em Class. Quantum Grav.}, 27:084024, 2010.

\bibitem{P070074}
S.~Hild et~al.
\newblock Photon pressure induced test mass deformation in gravitational‐wave
  detectors.
\newblock {\em Class. Quantum Grav.}, 24:5681‐5688, 2007.

\bibitem{P1100166}
H.~P. Daveloza et~al.
\newblock Controlling calibration errors in gravitational-wave detectors by
  precise location of calibration forces.
\newblock {\em Journal of Physics: Conference Series}, 363:012007, 2012.

\bibitem{P1500269}
B.~P. Abbott et~al.
\newblock {GW150914}: First results from the search for binary black hole
  coalescence with {A}dvanced {LIGO}.
\newblock 2016.

\bibitem{DarkhanAlog}
D.~Tuyenbayev.
\newblock {ETM} transmittivity measurement.
\newblock {\em {LHO} Log 13005}, July 25, 2014.

\bibitem{T1100068}
LIGO Photon~Calibrator Team.
\newblock Pcal final design document.
\newblock {\em LIGO Document Control Center}, T1100068, 2015.

\bibitem{T130442}
L.~Canete et~al.
\newblock Optical follower servo design for the calibration of a gravitational
  wave detector.
\newblock {\em LIGO Document Control Center}, T130442, 2013.

\bibitem{T1500131}
LIGO Photon~Calibrator Team.
\newblock {LHO} {Y-E}nd power sensor calibration trends.
\newblock {\em LIGO Document Control Center}, T1500131, 2015.

\bibitem{T1500036}
LIGO Photon~Calibrator Team.
\newblock Photon calibrator gold standard and checking standard {NIST}
  calibrations.
\newblock {\em LIGO Document Control Center}, T1500036, 2015.

\bibitem{T080316}
S.~Erickson.
\newblock Investigation of variations in the absolute calibration of the laser
  power sensors for the {LIGO} photon calibrators.
\newblock {\em LIGO Document Control Center}, T080316, 2008.

\bibitem{LLOalog}
S.~Kandhasamy et~al.
\newblock Frequency dependence of {P}cal calibration.
\newblock {\em {LLO} Log 18106}, May 13, 2015.

\bibitem{NIST-1297}
B.~N. Taylor and C.~E. Kuyatt.
\newblock Guidelines for evaluating and expressing the uncertainty of {NIST}
  measurement results.
\newblock {\em {NIST} Technical Document}, 1297, 1994.

\bibitem{G1501362}
J.~Driggers et~al.
\newblock Interferometer beam position.
\newblock {\em LIGO Document Control Center}, G1501362, 2015.

\bibitem{T1500377}
D.~Tuyenbayev et~al.
\newblock Improving {LIGO} calibration accuracy by tracking and compensating
  for slow temporal variations.
\newblock {\em LIGO Document Control Center}, P1600063, 2016.

\bibitem{timingiligo}
Y.~Aso et~al.
\newblock Accurate measurement of the time delay in the response of the {LIGO}
  gravitational wave detectors.
\newblock {\em Class. Quantum Grav.}, 26(5):055010, 2009.

\bibitem{HWINJ}
J.~Betzwieser et~al.
\newblock Documentation of the {A}dvanced {LIGO} hardware injection
  infrastructure.
\newblock {\em LIGO Document Control Center}, T1400349, 2015.

\bibitem{pcaldarmactuation}
R.~L. Savage and D.~Tuyanbayev.
\newblock Actuating the {DARM} loop using a {P}hoton calibrator.
\newblock {\em LIGO Document Control Center}, G1501352, 2015.

\end{thebibliography}
\clearpage
\end{document}